\pdfoutput=1
\documentclass[pdflatex,sn-mathphys-num]{sn-jnl}

\usepackage{geometry}
\usepackage{graphicx}%
\usepackage{multirow}%
\usepackage{amsmath,amssymb,amsfonts}%
\usepackage{amsthm}%
\usepackage{mathrsfs}%
\usepackage[title]{appendix}%
\usepackage{xcolor}%
\definecolor{b}{rgb}{0,0,1} 
\usepackage{textcomp}%
\usepackage{manyfoot}%
\usepackage{booktabs}%
\usepackage{subcaption}
\usepackage{parskip}
\usepackage{algorithm}%
\usepackage{algorithmicx}%
\usepackage{algpseudocode}%
\usepackage{listings}%
\usepackage{float}
\usepackage{siunitx} 
\usepackage{tabularx} 
\usepackage{array} 
\usepackage{longtable}
\usepackage{makecell} 
\usepackage{caption}
\usepackage{ltcaption}  
\usepackage{xltabular} 
\newlength{\simtablewidth}
\newlength{\decodingtablewidth}
\usepackage{helvet}

\usepackage{caption}
\DeclareCaptionFont{customeighthalf}{\fontsize{8.5pt}{11.2pt}\selectfont}
\DeclareCaptionLabelSeparator{pipe}{ \textbar{} }

\theoremstyle{thmstyleone}%
\theoremstyle{thmstyletwo}%

\theoremstyle{thmstylethree}%

\renewcommand{\thesubsection}{\Alph{subsection}.}
\begin{document}

\newgeometry{margin=12.7 mm} 

\title[HippocamPEX]{HippocamPEX: A Hippocampus-Inspired In-Silico Architecture for Context-Aware Sequence Prediction}

\author[1]{\fnm{Md Abdullah al} \sur{Kaiser}}\email{mkaiser8@wisc.edu}
\author[2,5]{\fnm{Keiland W.} \sur{Cooper}}\email{kwcooper@uci.edu}
\author[1]{\fnm{Rouhan} \sur{Noor}}\email{rnoor2@wisc.edu}
\equalcont{These authors contributed equally to this work.}
\author[1]{\fnm{Ankur} \sur{Singh}}\email{ankur.singh@wisc.edu}
\equalcont{These authors contributed equally to this work.}
\author[2,5]{\fnm{Zhishun} \sur{Yang}}\email{zhishuny@uci.edu}
\author[3]{\fnm{Hamed} \sur{Poursiami}}\email{hpoursia@gmu.edu}
\author[1]{\fnm{Bhuvanesh Sathish} \sur{Kayarabettu}}\email{kayarabettu@wisc.edu}
\author[4]{\fnm{Ziyi} \sur{Liang}}\email{liangz25@uci.edu}
\author[4,5]{\fnm{Babak } \sur{Shahbaba}}\email{babaks@uci.edu}
\author[3]{\fnm{Maryam } \sur{Parsa}}\email{mparsa@gmu.edu}
\author*[2,5]{\fnm{Norbert J.} \sur{Fortin}}\email{norbert.fortin@uci.edu}
\author*[1]{\fnm{Akhilesh} \sur{Jaiswal}}\email{akhilesh.jaiswal@wisc.edu}

\affil*[1]{\orgdiv{Department of Electrical and Computer Engineering}, \orgname{University of Wisconsin-Madison}, \state{WI}, \country{USA}}

\affil[2]{\orgdiv{Department of Neurobiology and Behavior}, \orgname{University of California Irvine}, \state{CA}, \country{USA}}

\affil[3]{\orgdiv{Department of Electrical and Computer Engineering}, \orgname{George Mason University}, \city{Fairfax}, \state{VA}, \country{USA}}

\affil[4]{\orgdiv{Department of Statistics}, \orgname{University of California Irvine}, \state{CA}, \country{USA}}

\affil[5]{\orgdiv{Center for the Neurobiology of Learning and Memory}, \orgname{University of California Irvine}, \state{CA}, \country{USA}}

\abstract {

The brain’s hippocampus enables episodic memory---the ability to encode events within their context, organize them into sequences, and generate predictive representations to guide adaptive behavior. Modern computing architectures, however, are optimized for large-scale arithmetic and struggle with such context-dependent sequential prediction or “cognitive” tasks. To translate these capabilities into a mixed-signal architecture, we first derived a working model from the neuroscience literature distilling key hippocampal computations into engineering-friendly abstractions. We then leveraged converging advances in in-memory computing, neuromorphic computing, asynchronous circuits, and analog computing to design a mixed-signal hardware architecture implementing these computations, called HippocamPEX. Finally, we demonstrated that HippocamPEX reproduces characteristic representational dynamics of hippocampal subregions and performs robust context-specific sequence prediction through detailed activity and circuit simulations. By tightly coupling hardware design with neuroscience knowledge, this work establishes a pathway for artificial systems capable of rapid, adaptive, and context-aware decision-making, addressing a central gap between biological and conventional computing.

}

\keywords{}

\maketitle

\newpage
\noindent
\normalsize


Modern computers have revolutionized a wide range of applications that require fast, large-scale arithmetic operations. However, today's computers  perform poorly on "cognitive" tasks: tasks that require learning from past experiences, and context-aware anticipation of future events. In contrast, the brain has evolved efficient solutions to such tasks through millions of years of tinkering and selection \cite{allen2013evolution}. A key brain region here is the hippocampus, a seahorse-shaped structure near the middle of the human brain that receives highly processed information from many other brain systems \cite{vanstrien2009aAnatomy, lavenex_hippocampal-neocortical_2000, furtak_functional_2007, kerr_functional_2007}. Considerable research has shown that the hippocampus is critical for episodic memory, our ability to remember specific sequences of events along with the context in which they occurred \cite{lisman1999relating, allen2013evolution, eichenbaum_medial_2007, stark2018s}. Underlying this ability is the hippocampus’ unique capacity to rapidly form event representations that are context-specific (event-in-context representations) \cite{eichenbaum_medial_2007, allen2016Nonspatiala, komorowski_robust_2009} and to temporally integrate them to capture the sequence of past, present, and future (predicted) events \cite{lisman1999relating, wikenheiser2015hippocampal, damphousse_vicarious_2026, levy1996sequence, shahbaba2022hippocampal}. These coding properties provide a framework by which our event memories are organized by common elements, which facilitates the subsequent recall of relevant past events in support of context-appropriate decision-making \cite{bornstein2017Reinstated, whittington_tolman-eichenbaum_2020, eichenbaum1999Hippocampus}. The same hippocampal properties also underlie our ability to build long-term knowledge (semantic memory), through the slow and gradual integration of recent event memories into existing knowledge structures in the cortex \cite{mcclelland_why_1995, kumaran_what_2016, teyler2007Hippocampal}. During decision-making, other brain structures, specifically the prefrontal cortex, are thought to support the process of evaluating memory-guided predictions from different brain systems, including event-based predictions from the hippocampus and knowledge-based predictions from the cortex, and of selecting the appropriate action \cite{euston2012Rolea, miller2001Integrativea, brown_neural_2026}. By incorporating such biologically motivated principles, hardware and software systems may leverage past experiences more effectively to support rapid adaptive decision-making in complex context-dependent and sequential tasks characteristic of biological intelligence.

Sequence learning and prediction have been studied in various neuromorphic and biologically inspired computing systems. Spaun performs basic pattern completion by generalizing from input sequences \cite{stewart2012spaun}, but it lacks the capability to predict future sequences from partial input cues, focusing instead on completing observed patterns. MemSpiking$^\text{TM}$ demonstrates context-aware sequence prediction of short, three-element sequences using a small 8×8 memristive crossbar array \cite{siegel2023demonstration}, while Resistive Random Access Memory (ReRAM) devices have also been employed as synaptic elements for sequence prediction, though such designs are typically context-independent \cite{bouhadjar2023sequence}. Intel’s Loihi chip implements a biologically inspired sequence learning algorithm with an offline hippocampal-like replay mechanism for memory consolidation and prediction generation \cite{lindsey2022sequence}, and the ROLLS neuromorphic processor leverages on-chip plasticity and winner-take-all dynamics to store, replay, and update sequences of states in real time \cite{kreiser2018organizing}. Recent work has further integrated state space models with spiking neural networks to achieve energy-efficient long-range sequence modeling, enabling deployment on neuromorphic hardware \cite{stan2024learning}. Despite these advances, existing approaches either perform sequence prediction independent of context, are limited to very short sequences, or rely on non-volatile devices that require significant CMOS process modifications. In other cases, they utilize full-scale general-purpose neuromorphic processors rather than brain-inspired application-specific hardware  based on architectural details of cognitive brain systems like the hippocampus.

Furthermore, the hippocampus’ crucial role in organizing memories has long inspired computational models aimed at enhancing artificial intelligence, particularly in reinforcement learning (RL). Early frameworks such as Model-Free Episodic Control (MFEC) \cite{blundell2016model} and Neural Episodic Control (NEC) \cite{pritzel2017neural} showed that caching high-value experiences can accelerate learning, but their dependence on similarity-based lookups over large, high-dimensional embeddings makes retrieval computationally costly as memory grows. Subsequent approaches addressed these limitations by combining predictive modeling with episodic recall. For example, MERLIN \cite{wayne2018unsupervised} integrates a generative predictive model with memory to capture long-range temporal dependencies in partially observable domains, while Model-Based Episodic Memory \cite{le2021model} unifies episodic recall with planning to support fast adaptation in stochastic environments. Building on this trajectory, we developed Hippocampal-Augmented Memory Integration (HAMI) \cite{poursiami2025scalable} to more closely align with key functional principles of the hippocampus. HAMI employs symbolic indexing and hierarchical refinement to encode events within their context and retrieve them efficiently, thereby facilitating adaptive decision-making in context-dependent sequential tasks.
This trajectory of algorithmic progress establishes a compelling motivation for creating specialized architectures that translate these brain-inspired memory functions into efficient, scalable hardware. 

\vspace{-3 mm} 
\begin{figure*}[h!]
   \centering
   \captionsetup{font=customeighthalf, labelsep=pipe}
   \includegraphics[width=0.95\linewidth]{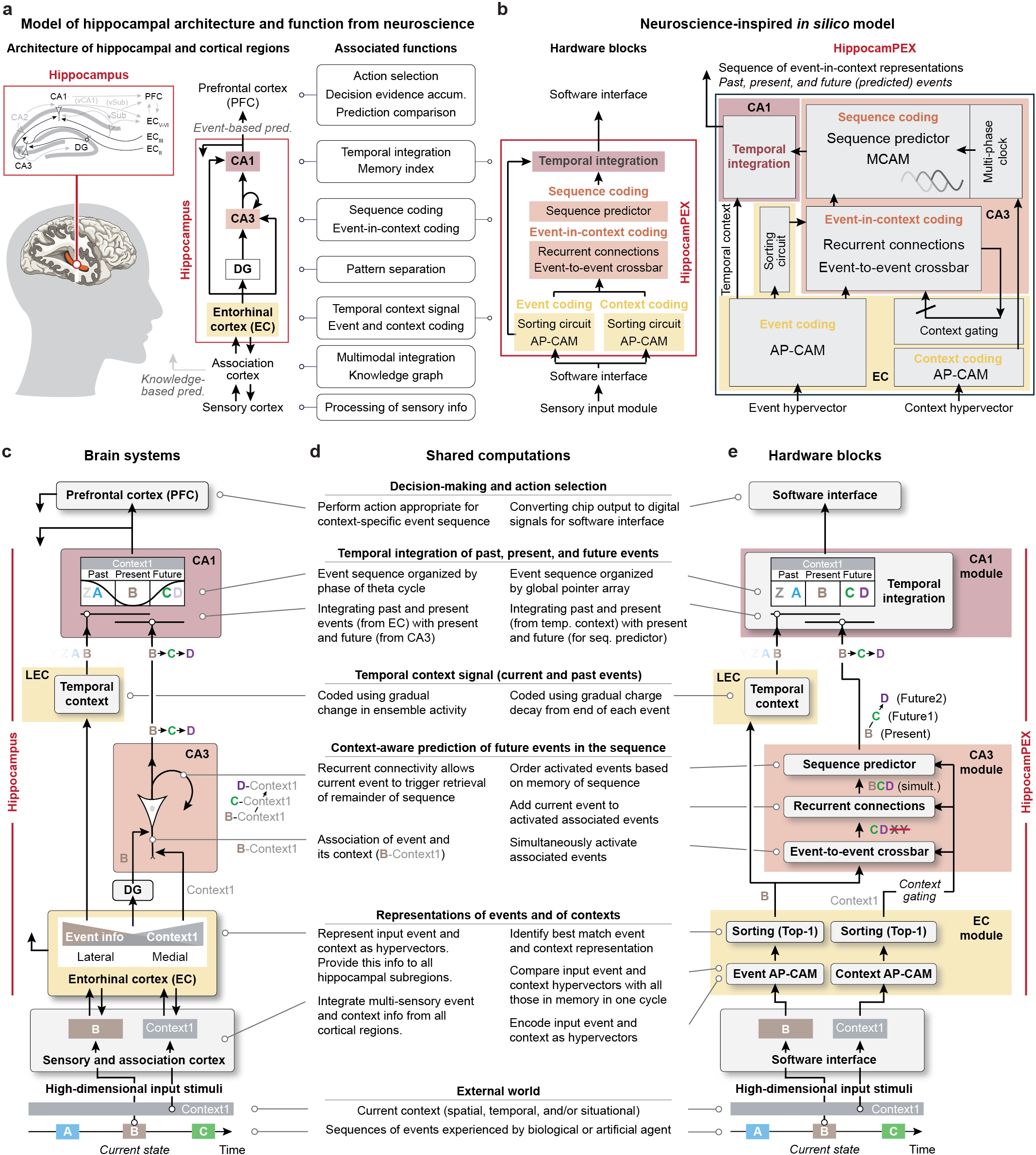}
   \vspace{2 mm}
   \caption{\selectfont \textbf{Overview of the brain architecture and functions abstracted in the HippocamPEX hardware.}
   (\textbf{a}) Working model of the anatomy of hippocampal and cortical regions and of their associated functions. To simplify the mapping with the hardware, here we consider the entorhinal cortex (EC) as part of the hippocampus and exclude the dentate gyrus (DG).
   (\textbf{b}) Corresponding engineering-friendly abstraction of key functions in the HippocamPEX architecture and hardware blocks. For clarity, individual hardware blocks (gray boxes) are grouped into modules that map onto hippocampal subregions (colored boxes).
   (\textbf{c-e}) Working model of corresponding flow of information and computations supporting context-dependent sequential tasks in the hippocampal system and HippocamPEX architecture.   
   Key hippocampal architecture (\textbf{c}) and subregional computations (\textbf{d}), with the corresponding architecture of hardware blocks in HippocamPEX (\textbf{e}).
   Note that we separated the temporal context function of the lateral entorhinal cortex (LEC) from the rest of EC to simplify the mapping with the hardware.
   }
\vspace{-5mm}
   \label{OverviewFig}
\end{figure*}

Here we address this critical gap by developing an end-to-end framework spanning neuroscience abstraction, hardware design, and computational validation to design a hippocampus-inspired mixed-signal architecture called HippocamPEX.
We introduce an engineering-friendly working model of the hippocampus implementable \textit{in silico}, which captures fundamental neuroscience principles including the functional segregation of event and context information in the cortex, their organization into sequences of event-in-context representations in the hippocampus to form episodic memories and support event-based predictions, and broader computational principles such as hierarchical neural representations and parallel information processing. 
We then leveraged converging advances in in-memory computing, neuromorphic computing, asynchronous circuits, and mixed-signal analog computing to design the HippocamPEX architecture, which implements key principles of the working model.
Subsequently, we introduce a software neuroscience model that simulates neural activity across hippocampal subregions during example event sequences, and quantifies the information contained in the ensemble activity of each subregion using neural decoding, which provides a platform to evaluate the degree of correspondence between subregional operations in the brain and in the \textit{in silico} hardware architecture. Finally, through detailed circuit simulations, we demonstrate that the HippocamPEX architecture mirrors the representational dynamics contained in simulated hippocampal subregional activity and successfully performs a context-specific sequence prediction task. 
The resulting HippocamPEX architecture lays foundational groundwork to map hierarchical and parallel processing of complex brain systems using state-of-the-art silicon chip design ecosystem. It also enables adaptive decision‑making in artificial systems closely inspired by the architecture of the biological brain, paving the way to solve complex problems similar to those solved by humans and other biological agents.

\vspace{3mm}
\section*{Results}

\vspace{-5 mm}
\noindent
\subsection*{\textit{Engineering-friendly Neuroscience Abstraction of Key Functions of the Hippocampal System}}
\normalsize

Our goal in this effort was \underline{not} to perfectly replicate hippocampal and cortical systems but to design hardware and software that capture the main principles by which these brain systems encode, retrieve, and organize memories as well as how they support memory-based decision-making and planning for the future. 
To achieve this, we developed a working model of the crucial neural architectures and computations  implementable \textit{in silico}, which we derived by integrating available evidence from the behavioral, neurophysiological, and computational neuroscience literature (Fig. \ref{OverviewFig}a,c). Note that we also focused on features that are novel to hardware architectures, can significantly improve performance and efficiency, and have the potential to subsequently scale up in size and complexity. 

Three functional principles of the brain serve as the foundation for our hardware design. 
The first is the hierarchical nature of the representations, whereby a hippocampal representation is linked to the corresponding information in entorhinal cortex (EC), association cortex, and sensory cortex \cite{teyler2007Hippocampal, mcclelland_why_1995}. This allows the hippocampus to operate on a smaller version (or index) of the full representation, which increases memory capacity and limits interference.
The second is the functional segregation of event and context information in the cortex, with their subsequent integration and sequencing in the hippocampus \cite{eichenbaum_medial_2007, knierim_functional_2014, allen2013evolution, lisman1999relating, stark2018s}. Contexts are generally defined as the background to an event presentation---the set of slowly changing features present at the time of the event that is shared with preceding and subsequent events \cite{stark2018s}. Note that the context often refers to a location in an environment or a specific room, but also extends to situations, internal states, and cognitive states \cite{polyn2008memory, maren_contextual_2013, stark2018s}. Through this organization, the hippocampus can rapidly encode an episodic memory (a specific sequence of event-in-context representations) and the cortex can maintain a stable representation of long-term knowledge.
The third is parallel information processing, specifically the notion that information is processed simultaneously across regions and that multiple systems will provide predictions for upcoming states or decisions. In particular, the hippocampus provides event-based predictions to the prefrontal cortex, whereas the association cortex provides knowledge-based predictions \cite{preston_interplay_2013, bar_proactive_2009}.
The following sections describe how these core hippocampal functions were mapped in hardware and software implementations, how the implemented computations corresponded to those observed in the hippocampal system, and how the resulting architecture performed on a representative hippocampus-dependent task. 
\subsection*{\textit{HippocamPEX: A Mixed-Signal Hippocampus-inspired Hardware Architecture}} 

The HippocamPEX hardware architecture we designed encapsulates core structural and functional attributes of the hippocampal system (Fig. \ref{OverviewFig}b,d,e) and contains distinct hardware blocks grouped into modules corresponding to its subregions. 
Like its counterpart in the brain \cite{vanstrien2009aAnatomy, knierim_functional_2014, eichenbaum_medial_2007}, HippocamPEX is modality-agnostic, as it receives highly processed sensory information, and can thus perform the same tasks on different types of sensory inputs (e.g., images, sonar, olfactory, or tactile information). 
In this implementation, sensory input is processed by a software interface, which provides the event and context inputs encoded as sparse \textit{n}-bit binary hypervectors. These hypervectors represent extracted features from sensory organs and can be obtained from various sensors using emerging in-sensor computing hardware paradigms \cite{yin2023iris, datta2022processing, kaiser2025voltage, ravichandran2026memristive}. These hypervectors, which represent the input from the cortex, then enter HippocamPEX via the EC module (entorhinal cortex).

\begin{figure*}[h]
    \centering
    \captionsetup{font=customeighthalf, labelsep=pipe}
    \includegraphics[width=.95\linewidth]{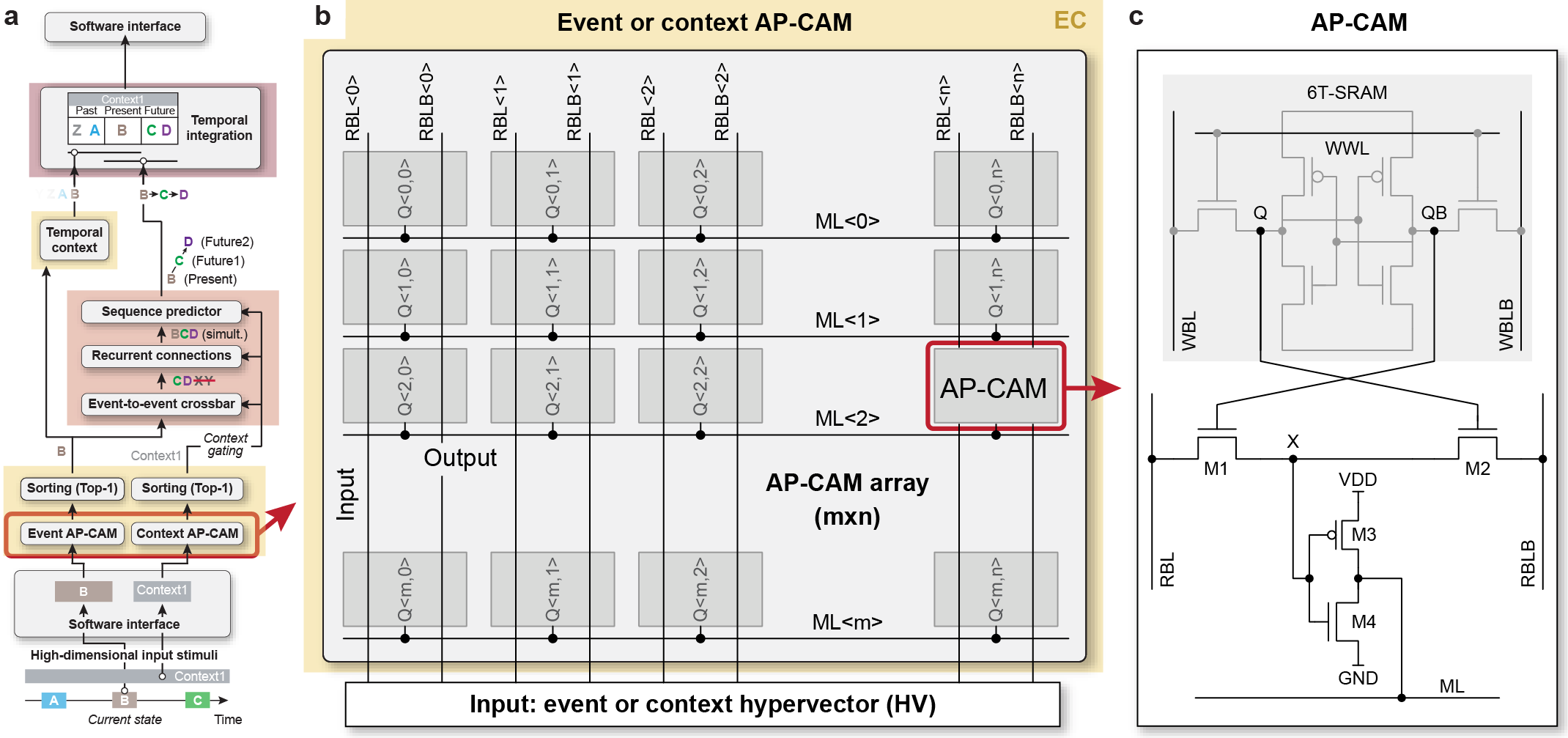} 
    \vspace{2 mm}
    \caption{\textbf{The entorhinal cortex (EC) module represents events and contexts as hypervectors using separate AP-CAM arrays.}
    (\textbf{a}) Location of the Event and Context AP-CAM blocks within the HippocamPEX architecture.
    (\textbf{b}) AP-CAM array (128 $\times$ 128) designed to store and match hypervectors (events or contexts). The input to the AP-CAM is the current event or context hypervector from the software interface, and the output is the voltage in the match lines (ML; rows) which reflects the degree of similarity between the current input hypervector and each stored hypervector.
    (\textbf{c}) Schematic of an example AP-CAM cell from the array showing the integrated 6T SRAM cell and a 4T CAM circuit.}
    \vspace{-5 mm}
    \label{fig_tmcam_array}
 \end{figure*}

In the EC module (see Methods), distinct analog proximity content-addressable memory (AP-CAM) blocks are used to encode event and context information (Event AP-CAM and Context AP-CAM; Fig. \ref{fig_tmcam_array}). The input event and context hypervectors then undergo an analog \textit{n}-bit similarity match operation with the stored event and context hypervectors. If there is no match, or if the match is below a certain threshold, the input will be encoded as a new event or context. If there is a match, the stored hypervectors that exhibit the highest number of matching bits with the input vector will be reflected in the degree of activation of the corresponding rows within the Event or Context AP-CAM. For example, if row 1 in the Context AP-CAM has the highest similarity with the input context hypervector, it signifies that the present context is most similar to previously encountered Context1. Similarly, if row 2 in the Event AP-CAM has the highest match, Event B is likely the current event. As a result, the representations of the event and context matching the current input are activated within the HippocamPEX hardware through the Event and Context AP-CAM blocks, which serve as the input to the CA3 module. 

\vspace{-2 mm}


\vspace{-2 mm}
\begin{figure*}[b!]
    \centering
    \captionsetup{font=customeighthalf, labelsep=pipe}
    \includegraphics[width=.95\linewidth]{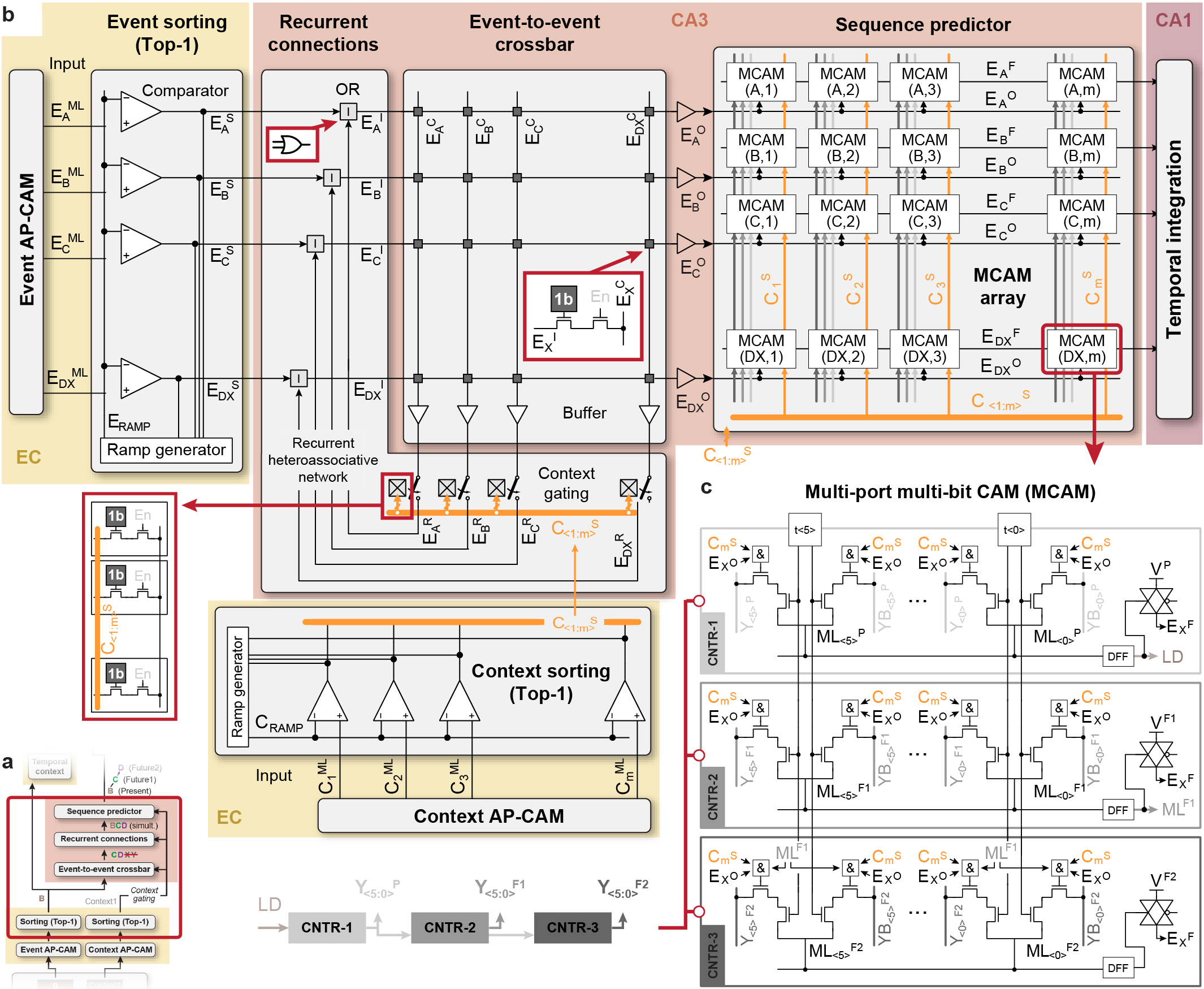}
    \vspace{2 mm}
    \caption{\textbf{The CA3 module generates context-aware predictions of the sequence of future events expected.}
    (\textbf{a}) Location of the corresponding blocks within the HippocamPEX architecture.
    (\textbf{b}) The inputs to the CA3 module are the representations of the present event and context obtained by identifying highest-matching event and context by the respective AP-CAMs through the event-sorting block. The event information enters the event-to-event crossbar block where previous associations among events are stored in the matrix of connections. 
    The recurrent connections block enables selective activation of connected events that were experienced in the present context (context-gating), and the sequence predictor block then orders those events by their time of occurrence.
    (\textbf{c}) Schematic of example 3-port, 6-bit CAM cell from the sequence predictor block with associated control circuits for applying predefined voltage tags corresponding to the present and two future events. Phase-shifted counters (\textit{left}) are used to encode the current event and the next two future (predicted) events. 
}
\vspace{-5 mm}    
    \label{fig_CA3module}
\end{figure*}    

In the CA3 module (see Methods), these matched event and context representations are then used to predict the sequence of future events within the given context (Fig. \ref{fig_CA3module}).
The first step of this prediction process involves the event-to-event (heteroassociative) crossbar block and the recurrent connections block, which together enable context-dependent connections between events (event-in-context coding). In our HippocamPEX hardware, 1-bit SRAMs are used to store those connections. For instance, if Event B occurs, the event-to-event crossbar may activate Events C, D, X, and Y, indicating these events were all previously associated (irrespective of the context). In this implementation, the recurrent connections block is used to add context-specificity to these event associations, by gating out activated events that are not relevant to the current context. To add to the previous example, if activated Events C and D were also experienced in Context1 (the current context in which Event B is experienced), their activation would be maintained; if Events X and Y were experienced in a different context, they would be filtered out. At this stage, HippocamPEX predicts the most probable future events in a given context, but their activation is simultaneous and thus provides no information about their order. 
The second step of the prediction process involves a sequence predictor block, which arranges the predicted events in an ordered sequence. Inspired by the phenomenon of phase precession in the hippocampus \cite{qasim_phase_2021, okeefe1993Phaseb, okeefe_dual_2005}, this block features a counter-based sequence prediction circuit that utilizes progressive multi-phase signals to compare temporal information of predicted events \textcolor{black}{stored in a multi-port CAM (MCAM) array. These multi-phase counters arrange the events in ascending temporal order as discrete analog voltages---future events are represented by elevated voltage levels relative to the present, with progressively higher voltages encoding events further into the future.} This information represents one of the two inputs to the CA1 module.

\begin{figure*}[h]
    \centering
    \captionsetup{font=customeighthalf, labelsep=pipe}
    \includegraphics[width=\linewidth]{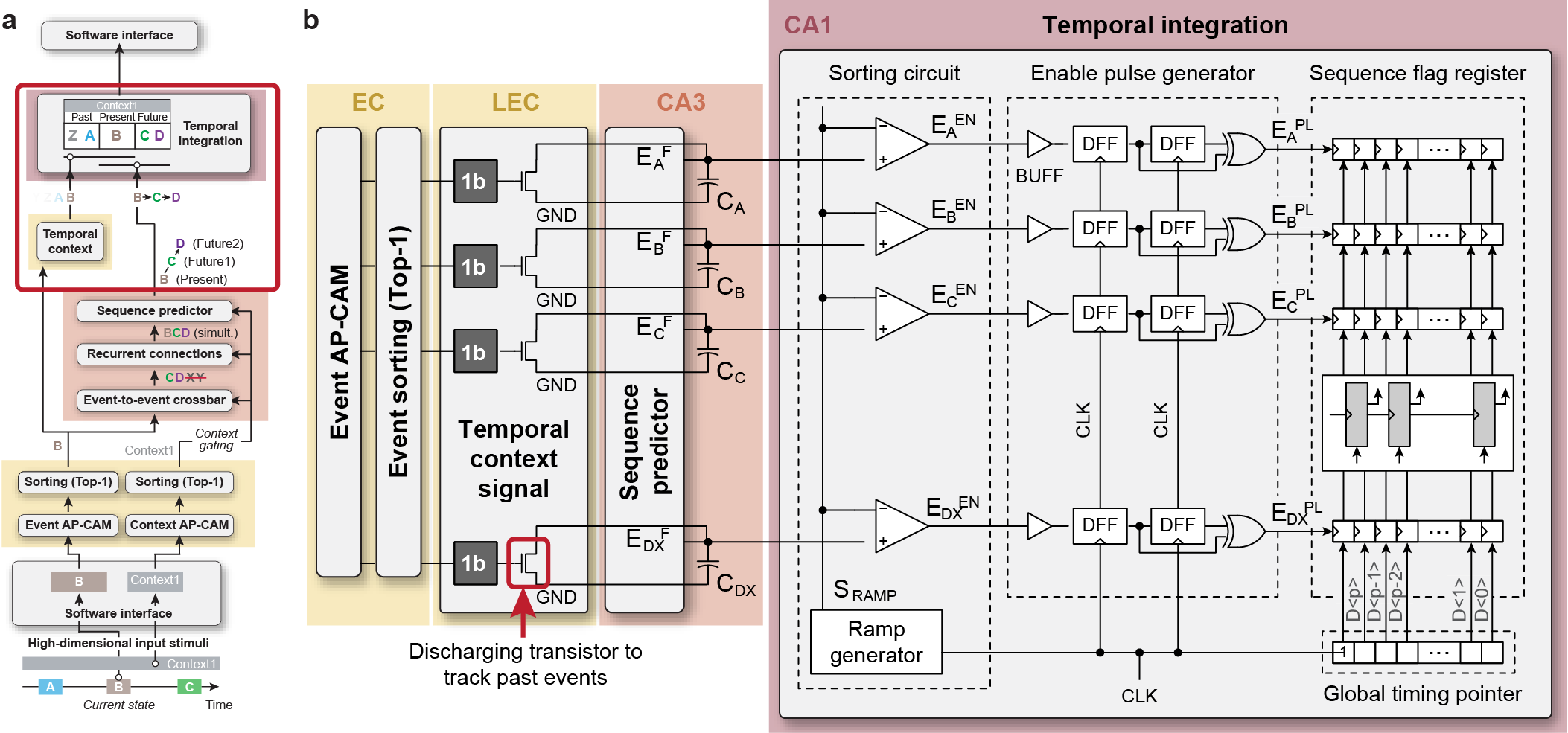}
    \vspace{-5 mm}
    \caption{\textbf{The CA1 module supports the temporal integration of past, present, and future events.}
    (\textbf{a}) Location of the corresponding blocks within the HippocamPEX architecture.
    (\textbf{b}) The temporal integration block is composed of a comparator-based sorting circuit, a pulse generator, sequence flag register, and global timing pointer. The inputs are the decaying representation of past events from the LEC module (temporal context signal) and the sequential activation of present and future events from the sequence predictor block of the CA3 module. The output is an integrated representation of past, present, and future events along with the corresponding global timing pointer value.    }
    \vspace{-10 mm}
    \label{fig_CA1}
\end{figure*}

The main function of the CA1 module (see Methods) is to integrate the sequence of past, present, and future events (Fig. \ref{fig_CA1}). 
As described above, the representations of present and future (predicted) events are provided by CA3's sequence predictor block. 
The representation of past events is provided by the temporal context signal block of the LEC module (lateral entorhinal cortex) \cite{tsao_integrating_2018, howard_distributed_2002, shankar_timing_2010}.
To track and retain past events in a sequence, the temporal context block employs a capacitor-based slow decay circuit to preserve historical information (see Fig. \ref{fig_CA1}b). For example, when the current event occurs (e.g., Event B), the corresponding electrical signal in the Event AP-CAM triggers the precharging of a capacitor linked to Event B. Similarly, if the preceding event was A, its associated capacitor was precharged at that time step. As a result, at any given moment, the capacitor of the most recent event holds the strongest signal, while past events exhibit progressively lower voltage levels due to charge decay. 
The temporal integration block of the CA1 module integrates this past, present, and future information by mapping the discrete analog voltages from the CA3 module (present and future events) and the analog capacitor discharging voltages from the LEC module (past events) to a digital code. This mapping is performed through an asynchronous latching operation of a register block (sequence flag register) based on enable pulses generated by progressive comparison with a ramp signal. The CA1 module's structured representation of past, present, and future events constitutes the output of HippocamPEX and reflects the event-based predictions outlined in the working model (Fig. \ref{OverviewFig}a,c). 


\vspace{-3 mm}


\noindent
\subsection*{\textit{The HippocamPEX architecture mimics key functional dynamics of hippocampal subregions}}
\normalsize

\begin{figure*}[b]
   \centering
   \captionsetup{font=customeighthalf, labelsep=pipe}
   \includegraphics[width=0.8\linewidth]{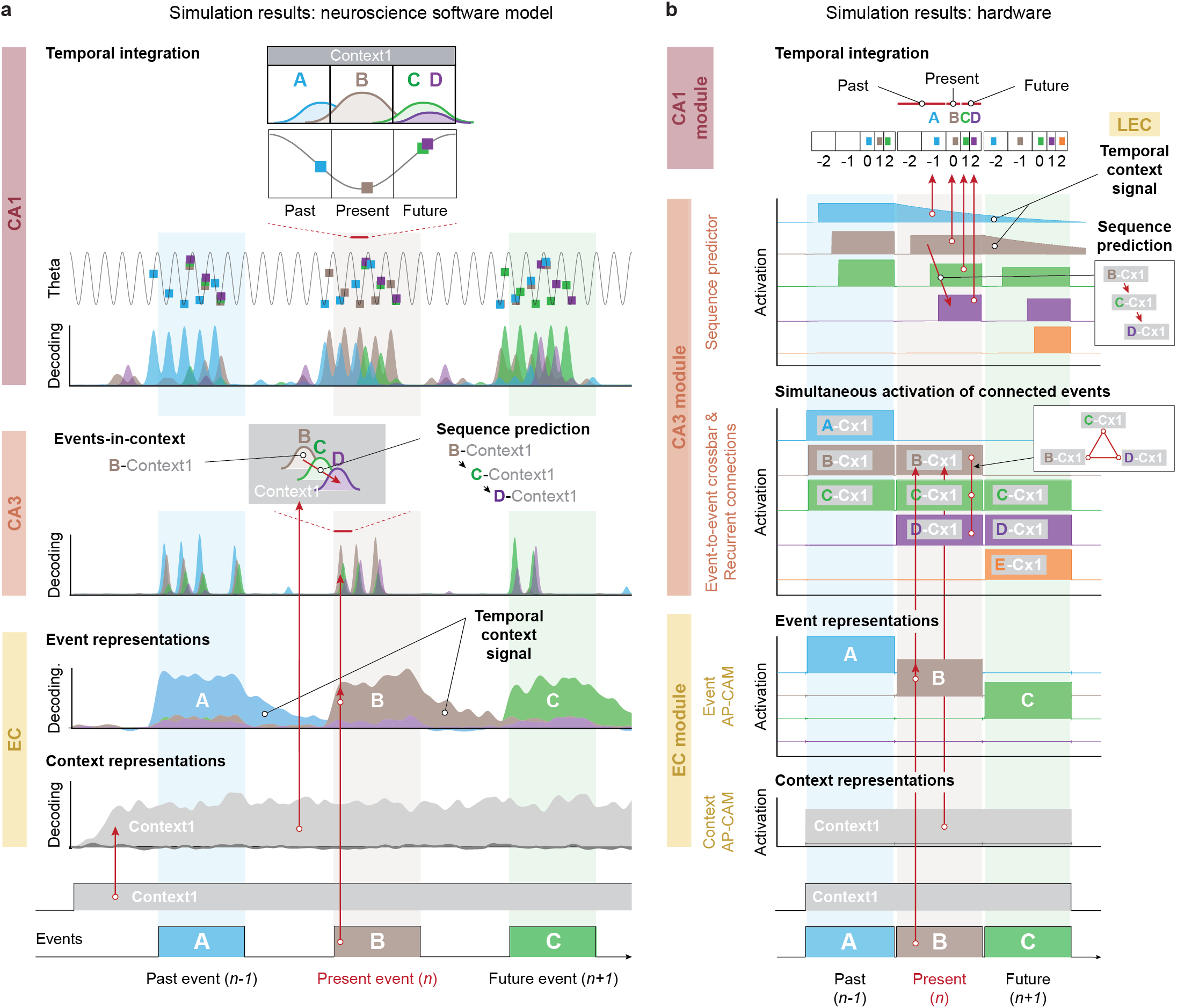}
    \vspace{1 mm}   
   \caption{\textbf{Functional agreement between the neuroscience software model and HippocamPEX hardware across hippocampal subregions.}
   (\textbf{a}) Neuroscience software model simulation output (neural decoding) across hippocampal subregions for an event sequence in a given context (Events A, B, and C in Context1; Event D not shown). The simulation replicates key neural representations recorded \textit{in vivo}. 
   (\textbf{b}) HippocamPEX hardware simulation results under matched conditions, demonstrating functional agreement in representations of events and of contexts (EC module), event-in-context representations and sequence prediction (CA3), temporal context signal (LEC), and temporal integration of past, present, and future events (CA1). Note that theta oscillations were not modeled in this implementation of the CA1 module.}
   \vspace{-3 mm} 
   \label{fig:SimulationResults} 
 \end{figure*}

After development of the HippocamPEX architecture, we proceeded to validate the computations of each module and their correspondence to those observed in the associated subregion of the brain. 
To achieve this, we formalized our neuroscience working model of Figure \ref{OverviewFig}c by developing a software model to more systematically examine the neural representations and computations in each subregion using simulations.
Specifically, we simulated neural activity across hippocampal subregions, and decoded the information contained in the ensemble activity in each subregion (see Methods). To build the software model, we first defined encoded firing rate functions that mirrored the expected encoding dynamics for each subregion based on our prior experiments and data from the literature \cite{tsao_integrating_2018, allen2016Nonspatiala, shahbaba2022hippocampal, komorowski_robust_2009}. To decode the content of neural activity in each subregion, we used a multinomial logistic regression model previously applied to a CA1 activity dataset recorded in a non-spatial sequence memory task \cite{shahbaba2022hippocampal}. 
Representative decodings of the simulated activity across subregions during an example episode (Events A, B, and C in Context1) are shown in Figure \ref{fig:SimulationResults}a, confirming that the software model captures core coding properties of hippocampal subregions.

The decodings of the software model were then compared to those obtained in a hardware simulation of the HippocamPEX architecture, which confirmed the correspondence in subregional computations (Fig. \ref{fig:SimulationResults}b).
The HippocamPEX circuits were implemented using the GlobalFoundries 22 nm FDX technology node and verified through Spectre simulations. To enable a direct comparison with the software model, the detailed SPICE-level simulation outputs were abstracted for visualization purposes, allowing functional similarities between the hardware and software representations to be more clearly demonstrated (see Methods). Hardware simulations were conducted under conditions that closely matched the software model, with the same context selected (Context1) and event sequence emulated (Events A, B, and C).
Consistent with the neuroscience software model, the hardware shows selective activation of the currently presented event in the EC module, which closely mirrors the sparse and event-specific activation observed in the software model. 
In the CA3 module, the hardware reproduces the event-in-context and sequence prediction representations by activating connected future events as a function of the current event and context. For example, when Event B is presented, the circuitry activates C and D as probable upcoming events (output of recurrent connection block; events filtered out by context not shown) before organizing them in the appropriate order (output of sequence predictor block). These pulse dynamics are consistent with the simulated CA3 coding of Fig. \ref{fig:SimulationResults}a, with the exception that the periodic repetition of the patterns associated with theta oscillations was not modeled in the hardware.  
Finally, in the CA1 module, the hardware demonstrates temporal integration that is functionally analogous to that observed in the software model.
In the hardware, counter values encode temporal relationships---the present event is mapped to a predetermined digital reference value, whereas past and future events are represented using monotonic digital codes, ensuring that the present event remains fixed at its designated reference level. 
Together, these results show that the HippocamPEX architecture faithfully mimics the functional dynamics of the hippocampal subregions captured by the neuroscience-based software model.

\begin{figure*}[h!]
   \centering
   \captionsetup{font=customeighthalf, labelsep=pipe}
   \includegraphics[width=.95\linewidth]{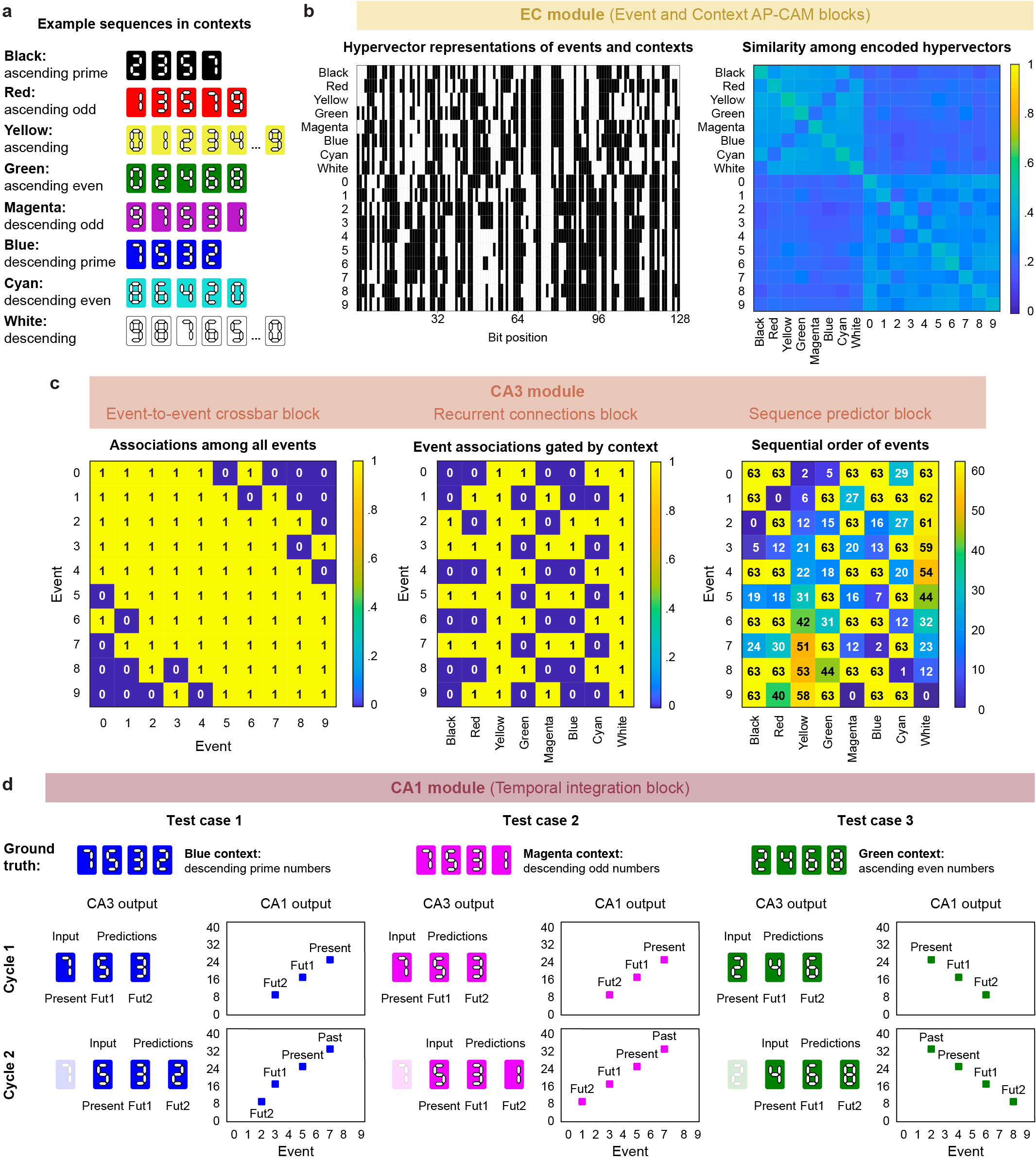}
   \vspace{1 mm}
   \caption{\textbf{HippocamPEX hardware model verification for context-aware sequence prediction task. }
   (\textbf{a}) Correct sequence of events (digits) for each context (colors). For clarity, the context-specific sequences are differentiated using digit-based rules (e.g., odd/even, prime/composite, ascending/descending). 
   (\textbf{b}) High-dimensional (128-bit) hypervector representations of events and contexts encoded in the AP-CAM blocks of the EC module (left) and cosine similarity across hypervectors (right). 
   (\textbf{c}) Context-gated event associations and sequential order representations in the CA3 module. The event-to-event crossbar block captures the connectivity of events in a sequence irrespective of context (1 = connected events; e.g., Event 2 is connected to all events except 9). The recurrent connections block maintains only associated events previously experienced in the present context (1 = active events; e.g., Event 2 in the Black Context only activates Events 3, 5, and 7). The sequence predictor block (MCAM array) encodes the order of each event in the sequence, with lower signal values indicating earlier positions in the sequence (the maximum value, 63, indicates events not included in the sequence). 
   (\textbf{d}) Three test cases demonstrating the integration of past, present, and future events in the CA1 module. On the first event presented (Cycle 1), CA1 receives the present event and two future (predicted) events from CA3 and no past event information from the temporal context block. On the second event (Cycle 2), CA3's output progresses by one sequence position and the temporal context block represents the previous digit as the past event (not shown). CA1 integrates past, present, and future events using counter values, with higher values indicating the past and lower values the future.
   }
   \vspace{-3 mm}
   \label{fig_hardware_model_verification}
 \end{figure*}

\vspace{-4 mm}
\noindent
\subsection*{\textit{The HippocamPEX architecture performs context-specific sequence prediction tasks }}

To assess the ability of the HippocamPEX architecture to support representative cognitive tasks, we evaluated its performance in a context-dependent sequence prediction paradigm. 
The challenge of this benchmark task is that, to respond correctly, the agent must remember the specific context and sequence position in which a given event occurs, which depends on the type of contextual and sequential representations specific to the hippocampal system.
In this task, digits serve as events and their background colors define contexts. Each color context enforces a different sequencing rule (e.g., odd/even, prime/composite, ascending/descending), so the correct continuation of a sequence cannot be inferred from the digit alone but requires consideration of its context. 
By using digit sequences embedded in color contexts, a design inspired by standard paradigms in human memory experiments \cite{stark2018s}, the task provides a controlled yet biologically relevant benchmark for testing context-aware sequence learning and prediction.



In this implementation, we encoded events and contexts using high-dimensional binary vectors that capture structured similarity. Context vectors were built compositionally from RGB bases, and digit vectors were generated from seven-segment encodings. This design ensures that related colors (e.g., cyan and blue) or digits (e.g., 8 and 9) share correlations in their hypervector representation, unlike orthogonal encodings that treat all items as equally distant. 
Incorporating such correlations makes the task more realistic and challenging: the system must distinguish overlapping representations while remaining robust to noise or perturbations, a property central to hippocampal memory functions and essential for evaluating HippocamPEX. 
The HippocamPEX hardware model began with an off-line training phase, where event and context hypervectors, the association matrix, and temporal information were generated. These trained parameters were then configured into the memory blocks (AP-CAM, crossbar, MCAM) during the initialization phase, with future work aiming to integrate training fully on-chip. The event and context hypervectors were loaded into their respective AP-CAMs of the EC module, while the association matrix was mapped onto the event-to-event crossbar and recurrent connections blocks, and the order was stored in the MCAM array of the sequence predictor block of the CA3 module. 
The detailed inference procedures including top-1 sorting, counter initialization, voltage tagging, and temporal context signal generation are described in the Methods (see Hardware Circuit Implementation) and in \textit{Algorithm 1} (see Hardware model and verification). 

Our results show that the HippocamPEX architecture accurately extracted the structure of the relationships across events and contexts experienced, as well as predicted and integrated the appropriate sequence of events in each context (Fig. \ref{fig_hardware_model_verification}). 
First, the EC module (AP-CAMs) encoded the event and context hypervectors and captured their inherent similarities (Fig. \ref{fig_hardware_model_verification}b). 
Second, the different blocks in the CA3 module performed the expected operations to yield context-specific sequential predictions (Fig. \ref{fig_hardware_model_verification}c). More precisely, the event-to-event crossbar represented associations among events irrespective of context, the recurrent connections filtered out retrieved events not experienced in the present context (context-gating), and the sequence predictor preserved their sequential order within a context.
Third, the CA1 module integrated the appropriate past, present, and future events for each context-specific sequence of events (Fig. \ref{fig_hardware_model_verification}d).
Collectively, these findings demonstrate that the HippocamPEX hardware successfully performs context-specific sequence prediction tasks, a hallmark of hippocampal function. 

\section*{Discussion}
\normalsize

This article introduced HippocamPEX, a hardware architecture inspired by the hippocampal system of the brain. 
We began by developing a working model of the key anatomical and functional elements by which the hippocampus supports complex cognitive tasks, in this case contextual and sequential decision-making, which were then mapped in hardware and software implementations. 
In the resulting HippocamPEX architecture, the EC module (entorhinal cortex) represents events and contexts as hypervectors using AP-CAM blocks, the CA3 module generates context-aware predictions of expected future events using three distinct blocks (event-to-event crossbar, recurrent connections, and sequence predictor), and the CA1 module supports the temporal integration of past, present, and future events using converging inputs from the temporal context block of the LEC module (lateral entorhinal cortex) and the CA3 module.
We then demonstrated that the HippocamPEX architecture mirrored the representational dynamics contained in simulated hippocampal subregional activity and successfully performed a context-specific sequence prediction task. More specifically, the HippocamPEX architecture accurately extracted the structure of the relationships across events and contexts experienced, as well as predicted and integrated the appropriate sequence of events in each context.
Collectively, these findings demonstrate that the HippocamPEX hardware effectively leverages the unique architecture, representational dynamics, and functionality of hippocampal subregions to solve the type of complex cognitive tasks supported by the hippocampus. 

This novel architecture of HippocamPEX offers several advantages. 
First, the mixed-signal hardware architecture leverages modern techniques including compute-in-memory (as represented by AP-CAM, event-to-event crossbar, MCAM) and tightly integrated digital-analog circuits. Unlike the traditional von Neumann architecture of modern computers, this new architecture avoids the energy-expensive transfer of event and context hypervectors across blocks. Instead, like memory indices in the brain, it relies on symbolic pointers to those hypervector representations which enable rapid and efficient parallel processing of the information across modules. 
Second, like the hippocampus itself, this architecture is modality-agnostic which effectively makes it broadly applicable for solving context-specific and sequential problems. In other words, the hippocampus is known to essentially perform the same operations across mammals using inputs specific to each species. These inputs can vary considerably across species, including differences in input type (e.g., primarily visual in primates, but sonar-like in bats) and complexity (e.g., more abstracted visual representations in primates than in rodents). The framework of event and context representations also applies across species and modalities; in fact, events tend to be defined as the unique conjunction of stimuli (of any modality) occurring at a discrete moment, and contexts as the set of external and internal cues or attributes that persists across events. The same logic applies to HippocamPEX---it is designed to process and integrate event and context information from any type of sensor input and is thus inherently flexible in its applications. 

Furthermore, the architectural framework of HippocamPEX is device-agnostic. Although our current implementation utilizes CMOS technology, it can be seamlessly adapted to emerging electronic \cite{aguirre2024hardware, yan2024emerging} and photonic technologies \cite{wang2025integrated, shen2023deep}. The compute-in-memory paradigm also naturally aligns with characteristics of memristive devices, phase-change memory (PCM), resistive RAM (ReRAM), and other emerging nanoscale electronic technologies that offer potential advantages including higher density, lower power consumption, and intrinsic analog computation capabilities \cite{zhong2021dynamic,zhong2022memristor}.
Finally, the principle of compute-in-memory combined with symbolic representation is scalable. As such, our approach can extend beyond the current number of events and contexts, as well as beyond the hippocampus to expand into increasingly complex neural architectures capturing spatial and temporal processing characteristics across cortical and subcortical structures.

Future efforts will incorporate functionalities not implemented in this first iteration of the HippocamPEX architecture. In terms of module-specific functions, this includes the addition of a dentate gyrus module to implement pattern separation \cite{borzello_assessments_2023}, which reduces interference between similar inputs by orthogonalizing their representations, and a CA3 module that more closely matches the operations reported \textit{in vivo}. More precisely, in this implementation, CA3’s recurrent connections were used for context-gating, whereas, in the brain, recurrent connections are primarily associated with pattern completion (the ability to reactivate a full pattern based on partial input) \cite{guzman_synaptic_2016, rolls2013Mechanismsa, neunuebel2014ca3} and sequence coding (the function served by our sequence predictor block) \cite{johnson2007Neurala, levy1996sequence, lisman1999relating, buzsaki2018space,damphousse_vicarious_2026}. Potential additions to the CA1 module include theta oscillations during online behavior \cite{johnson2007Neurala, wikenheiser2015hippocampal, shahbaba2022hippocampal}, as well as a replay function during offline periods (i.e., brief rest or sleep) \cite{carr_hippocampal_2011, foster_replay_2017} during which recent memories may be sporadically played back to the cortex to slowly integrate them into its semantic knowledge structure without disrupting older memories \cite{mcclelland_why_1995, kumaran_what_2016}. 
Beyond specific modules, future work can also focus on broader implementation strategies that more closely reflect information processing in the biological brain. This can include implementing statistical approaches to quantify uncertainty at each information processing step, to parallel the ability of humans to assess their confidence level before making decisions. It can also include extending the proposed architecture to support spiking neural network implementations, to mirror neural communication in the brain. While the present design operates on symbolic hypervector matching rather than discrete spike events, the general framework is compatible with event-driven spiking representations through software-hardware co-design.

\newpage
\section*{Methods}


\subsection{Hardware Circuit Implementation} 


\textbf{AP-CAM block (EC module).} 
Two separate AP-CAM arrays store events and contexts, respectively, encoded as hypervectors (HVs) in an \textit{n}-bit (\textit{n} = 128) binary format as rows of six transistor (6T) SRAM cell arrays of size \textit{m}\si{\times}\textit{n} (as shown in Fig. \ref{fig_tmcam_array}b), where \textit{m} represents the maximum number of events (or contexts) supported by the hardware. To enable efficient similarity match between stored (previously encountered) event and context hypervectors with that of the current input event and context hypervector, each 6T SRAM cell is integrated with a 4T Content-Addressable Memory (CAM) circuit 
capable of performing massively parallel in-memory search operations. The CAM compares all stored \textit{n}-bit patterns---i.e., previously learned event HVs in the Event AP-CAM or context HVs in the Context AP-CAM---against the input \textit{n}-bit query (event or context, respectively). When a perfect match occurs, the corresponding match line (ML) of the row remains at a high voltage level (VDD). However, upon detecting mismatches, the voltage on ML drops in a controlled way, enabling the degree of similarity (or proximity of the present query HV with the stored HVs) to be expressed as an analog value rather than a purely binary outcome \cite{yin2020xnor,zhang2024semantic}. 

Within each cell of the AP-CAM array (Fig. \ref{fig_tmcam_array}c), the write mechanism remains the same as in a conventional 6T-SRAM, utilizing write wordlines (WWL) and complementary write bitlines (WBL, WBLB), while dedicated read bitlines (RBL and RBLB) are used for search operations. 
A key feature of this design is the use of an in-situ XOR operation, implemented through transistors M1 and M2, which perform bitwise comparison between the stored data (Q and QB) and the input data lines (RBL and RBLB). When the stored and input bits match—such as Q = `0' with RBL = `0'—transistor M2 stays OFF while M1 turns ON, pulling node X to `0' via M1 and RBL. Similarly, if Q = `1' and RBL = `1', node X is pulled to `0' through M2 and RBLB. In contrast, when there is a mismatch—say Q = `0' and RBL = `1'—M1 conducts due to QB = `1', setting X to `1'. Likewise, for the case where Q = `1' and RBL = `0', X becomes `1' through M2 because RBLB = `1'. 

Transistors M3 and M4, connected between node X and ML, establish a controlled charging or discharging path that depends on the voltage at X, which reflects the bit-level similarity of query HVs with the stored HVs in the AP-CAM. Specifically, when a stored bit matches the input, node X is pulled low (`0'), turning on M3 and effectively connecting ML to VDD through M3. Conversely, in the case of a mismatch, node X rises high (`1'), activating M4 and thereby linking ML to GND via M4. The sizes (drive strengths) of transistors M3 and M4 directly determine the effective resistance between ML and the supply rails. Note that we are only interested in identifying if the input query (event or context) HV is similar to stored HVs, that is if the bits in query HV exhibit 90\% or higher matches with the bits in stored HVs. If none of the HVs stored in the AP-CAM exhibit a match of 90\% or greater, it implies that the hardware has not previously encountered events or contexts that sufficiently resemble the present ones.  

We leverage the fact that the hardware only needs to identify 90\% or above match (i.e., 10\% or less mismatch) between query and stored HVs to our advantage, since 10\% mismatch corresponds to roughly 13-bit mismatch among 128 bits. Thus, only 13 levels of voltages are needed to be represented on the analog ML line instead of 128 levels, significantly increasing the voltage resolution.
To ensure the circuit precisely differentiates up to 13-bit mismatches for a 128-bit HV, we carefully configured transistor strengths: the pull-down transistor M4 was designed stronger than the pull-up M3. This intentional asymmetry means that for a single bit mismatch among 128 bits, the path from ML to GND exhibits a lower resistance than the pull-up path to VDD, resulting in a larger voltage drop per mismatch. Such a design choice enhances the voltage step resolution, making the circuit more robust against the inevitable local variations and mismatch effects in silicon. Overall, this circuit technique enables our AP-CAM to maintain an accurate analog proximity similarity index, even in the presence of process variations.

\textbf{Top-1 sorting circuit (EC module).} The Event AP-CAM and Context AP-CAM arrays described above generate analog output voltages on the match lines for each event and context reflecting their degree of overlap with the present input. Following the match operation, a sorting block executes a top-1 search, selecting only the event and context with the highest analog voltage, which signifies the closest match. The top-1 sorting circuit comprises comparator blocks that evaluate the analog voltage of each match line in the Event AP-CAM (i.e., \si{E_A^{ML}}, \si{E_B^{ML}}, \si{E_C^{ML}}, …, \si{E_{DX}^{ML}} as shown in Fig. \ref{fig_CA3module}b) and Context AP-CAM (i.e., \si{C_1^{ML}}, \si{C_2^{ML}}, \si{C_3^{ML}}, …, \si{C_m^{ML}} as shown in Fig. \ref{fig_CA3module}b) against a ramp voltage (\si{E_{RAMP}} and \si{C_{RAMP}}, respectively). This common ramp signal, connected to the negative terminal of the comparator, gradually decreases from VDD to GND. When a comparator generates a high signal, it signifies that the ramp signal has intersected with the strongest matched analog voltage signal. This prompts the ramp generator to stop the ramp and halt the comparator's clock, thereby terminating further comparisons. Through this approach, the circuit efficiently conducts top-1 sorting, determining the most strongly matched event or context within the Event or Context AP-CAM, respectively. The sorted event (i.e., \si{E_A^S}, \si{E_B^S}, \si{E_C^S}, …, \si{E_{DX}^S} as shown in Fig. \ref{fig_CA3module}b) and context (i.e., \si{C_1^S}, \si{C_2^S}, \si{C_3^S}, …, \si{C_m^S} as shown in Fig. \ref{fig_CA3module}b) signals are one-hot encoded, ensuring that activation occurs only on the line corresponding to the event or context exhibiting the highest match with the present input. These top-1 sorted signals are then forwarded to the heteroassociative crossbar network. Note that instead of restricting the operation to top-1 sorting, we can extend it to top-k sorting, enabling more probabilistic and complex computations in CA3. This approach will be explored further in our future work.

\textbf{Event-to-event association crossbar (CA3 module).} The association network comprises a 128 \si{\times} 128 crossbar array that captures event-to-event associations (Fig. \ref{fig_CA3module}b). 
 Each of the row lines (i.e., \si{E_A^I}, \si{E_B^I}, \si{E_C^I}, …, \si{E_{DX}^I}) and column lines (i.e., \si{E_A^C}, \si{E_B^C}, \si{E_C^C}, …, \si{E_{DX}^C}) represents 128 distinct events (i.e., events A, B, C, …, DX). At each intersection, an SRAM-augmented AND gate can activate the column line based on the row line's activation and the previously learned associations between any two events. For instance, if Event B is connected to Events C, D, X, and Y, the SRAM at the intersections of those events will store a binary `1' in its memory during the learning phase of HippocamPEX. Consequently, in an example sequence of ZABCD, when the current input is similar to Event B, the association network will activate Events C, D, X, and Y. The activated events undergo context-gating as described below. Note that, in this hardware implementation, we have accounted for 128 events and `m' contexts (for example, m = 10). However, the architecture remains scalable, allowing for an arbitrary number of events and contexts within the HippocamPEX hardware.

\textbf{Context-gating circuit (CA3 module).} At the edges of the association crossbar, a context-gating block is connected to each event (Fig. \ref{fig_CA3module}b). This context-gating block filters the activated events based on the current input context signal. The gating switch will only be activated if the event is relevant to the present context. After activating all events connected to the input event by the association network, this circuit filters out those that are not relevant to the current context. Each event line is linked to a \si{1\times m} SRAM-augmented AND-OR operation, where `m' represents the number of contexts considered in the hardware. Each SRAM stores the association information for whether a specific event is linked to a particular context. For example, when the present input is Event B, Events C, D, X, and Y are being activated by the association network. If Events C and D are associated with Context1, while Events X and Y are associated with Context2, then, if the current input is Context1, the context line (\si{C_1^S}) will be activated from the context AP-CAM and the top-1 sorting block. When the context gating SRAM stores a `1’, the context activation line sends a control signal to enable the switch, activating it. As a result, events linked to that specific context will be passed through the recurrent connections (\si{E_B^C} to \si{E_B^R} and \si{E_C^C} to \si{E_B^R}), while irrelevant events are gated out. In this example, Events X and Y will not be transmitted, even though their column lines are activated in the event-to-event crossbar.

The activated events after the context-gating block are transmitted to the row event lines through a recurrent connection, utilizing an OR operation between the recurrent (e.g., \si{E_C^R}) and forward (e.g., \si{E_C^S}) event signals (originating from the Event AP-CAM and top-1 sorting block). This recurrent connection ensures the activation of row lines in the event-to-event crossbar corresponding to the present event and its associated context filtered events. For example, based on the previous example, Events C and D will be activated alongside the present input Event B. As a result, the row lines \si{E_B^O}, \si{E_C^O}, and \si{E_D^O} when the present event is Event B are triggered and sent to the sequence predictor block. This recurrent network guarantees that both the present event and the context-aware connected events are activated and passed on to the next stage.

\textbf{Sequence predictor block (CA3 module).} In the sequence predictor block (Fig. \ref{fig_CA3module}b), columns representing contexts are controlled by the sorted context (\si{C_1^S}, \si{C_2^S}, \si{C_3^S}, …, \si{C_m^S}) from the context sorting block, while the rows are driven by the event-in-context association signals (\si{E_B^O}, \si{E_C^O}, and \si{E_D^O}). A multi-bit SRAM at each intersection of rows and columns holds the corresponding temporal information about when each event occurred in the specific context.  To enable sequence prediction, a 3-port CAM is implemented (as shown in Fig. \ref{fig_CA3module}c), where each SRAM cell (labeled as \si{t<0:5>}) is augmented with additional circuit allowing the system to identify one present and predict two future events based on the temporal associations already stored in the SRAM cells. Note that while each MCAM integrates a 6-bit SRAM to store timing information, this temporal resolution is scalable, meaning higher-order memory per MCAM can be employed if necessary to enhance timing precision for different context-event combinations. 

The top-1 sorted context from the context AP-CAM block, together with the present event and the context-aware connected events, are activated in the recurrent crossbar network. Following these activations, CNTR-1 is enabled to establish the association between the present event and context. For example, if the present context is Context1, \si{C_1^S} will be activated, while all other context signals (\si{C_p^S}, where p = 2 \text{ to } m) remain inactive. Along the activated rows, as the present event and its connected events are enabled, CNTR-1 performs the matching operation by comparing the counter values (\si{Y_{<5:0>}^P}) only against those specific events on that particular column (context). Owing to the temporal assignment stored in the MCAM, the present event retains the lowest counter value relative to its next two future events. Consequently, the present event matches with CNTR-1 first, followed sequentially by the prediction of future events. When the CNTR-1 values (\si{Y_{<5:0>}^P}) perfectly match the stored values (\si{t<5:0>}) in the designated MCAM cell, the match lines (i.e., \si{ML_{<5>}^P, …, ML_{<0>}^P}) of port-1—dedicated to comparing the CNTR-1 values with the temporal information of the present context-event pair—will be set to `1’ in each MCAM cell. Consequently, a perfect match will generate a load (LD) signal. This counter value is then used to initialize CNTR-2 and CNTR-3 as an offset to their initial phase, which are responsible for predicting the next two events. Additionally, port-1 of the MCAM block precharges the event line (\si{E_B^F}, considering Event B is the present input in the sequence ZABCD) to a specific voltage (\si{V^P}), marking the event as the current active event.

In the subsequent phase, CNTR-2 and CNTR-3 are activated to predict the next two future events. For example, continuing from the previous scenario, Events C and D are the most probable occurrences following Event B within Context1. As a result, the MCAM cells corresponding to Events C and D in Context1 (i.e., row 3 and row 4 in column 1 of the MCAM array) become active. CNTR-2 values (\si{Y_{<5:0>}^{F1}}) then compare the stored temporal information of these activated events to identify an exact match. The event with a perfect match in CNTR-2 signifies the immediate next event in the sequence. Similarly, CNTR-3 searches for the subsequent event that follows the immediate future event. Additionally, CNTR-2 and CNTR-3 precharge the respective event lines to voltages \si{V^{F1}} and \si{V^{F2}} (Fig. \ref{fig_CA3module}c), indicating the immediate and next future events, respectively.

\textbf{Temporal context signal block (LEC module).} The temporal context signal block models the decaying trace of past events using a capacitor discharging circuit (Fig. \ref{fig_CA1}b). This block comprises a holding capacitor (i.e., \si{C_A}, \si{C_B}, \si{C_C}, ..., \si{C_{DX}}) to store the timing information and an SRAM-augmented discharging transistor for each event line. When a specific event is the present input, its corresponding capacitor is precharged to a voltage level of \si{V^P}. Initially, all SRAM bitcells are reset to `0'. Based on the detected present input event, the Event AP-CAM and top-1 sorting block update the relevant SRAM bitcell, triggering a gradual discharge of the holding capacitor. As the sequence progresses, the level of discharge increases, representing the fading memory of past events.
For example, consider the event sequence ZAB, where B is the present event, A is the recent past event, and Z is a more distant past event. When Event Z was the current input, its holding capacitor (\si{C_Z}) was precharged to \si{V^P} using the MCAM block in the sequence predictor, and the corresponding SRAM cell was set to `1’, initiating a slow discharge. By the time Event A becomes the present input (Event Z becomes the near past event), \si{C_Z} will have been discharged by a certain voltage drop (e.g., \si{V_{DROP}}) as a function of time. At this point, \si{C_A} is precharged to \si{V^P}, while \si{C_Z} retains a reduced charge of \si{V^P - V_{DROP1}}. When the sequence advances to Event B, the capacitors hold the following charges: \si{C_B} at \si{V^P}, \si{C_A} at \si{V^P - V_{DROP2}}, and \si{C_Z} at \si{V^P - V_{DROP3}}. In summary, the charge stored on the present event line will be at its maximum, while the voltage on the distant past event line will have decayed to a lower level (a function of the sequence length and timing between the events). In contrast, the near past event will retain a charge level between that of the present and the distant past event lines. Through this capacitively controlled discharging mechanism, the HippocamPEX architecture effectively maintains traces of past events.

\textbf{Temporal integration block (CA1 module).} 
The temporal integration block mirrors the functionality of the CA1 region in the hippocampus (Fig. \ref{fig_CA1}b). Each event line is linked to a holding capacitor (i.e., \si{C_A}, \si{C_B}, \si{C_C}, ..., \si{C_{DX}}). The temporal context signal block controls the discharge of past events, resulting in lower voltage levels on their respective capacitors compared to the present event voltage (\si{V^P}). Additionally, the MCAM block within the sequence predictor circuit precharges the capacitors corresponding to the next two predicted events to voltages \si{V^{F1}} and \si{V^{F2}}, where \si{V^{F2} > V^{F1} > V^P}. By sorting the voltage levels on the holding capacitors of all event lines, the CA1 block generates a sequence that integrates past, present, and predicted future events.
This is achieved using a comparator-based sorting circuit, a pulse generator that produces a pulse upon comparator switching, a group of sequence flag registers that record past, present, and future events in a sequence and a global timing pointer. For instance, in an event sequence ZABCD within a specific context, where Event B is the current input, the corresponding capacitor voltages are \si{C_Z} = \si{V^P - V_{DROP3}}, \si{C_A} = \si{V^P - V_{DROP2}}, \si{C_B} = \si{V^P}, \si{C_C} = \si{V^{F1}}, and \si{C_D} = \si{V^{F2}}. During the sorting operation, the node voltages of these capacitors are compared against a gradually decreasing ramp signal starting from VDD (where \si{V^{F2} < VDD}). The comparator linked to Event D detects when the ramp signal falls below \si{V^{F2}}, generating an activation signal (\si{E_D^{EN}}). This signal is then processed by the pulse generator—comprising two D Flip-Flops (DFFs) and an XOR gate—to produce a pulse signal (\si{E_D^{PL}}), which serves as the clock input for the sequencer flag register of Event D.

A free-running binary counter is used as a global timing pointer to integrate past, present, and future events in the sequence. The counter’s output is connected to the data inputs of the sequencer's flag registers. When a specific flag register corresponding to an event (e.g., \si{E_D^{PL}} for Event D) receives a clock pulse, it latches the current value of the timing counter. The system continues to sort events, generating activation pulses sequentially for Events C, B, A, and Z. As a result, each event’s flag register stores a counter value that reflects its position in the sequence. In this scheme, pulses generated earlier correspond to future events, while pulses generated later represent past events. Consequently, lower counter values indicate future events, and higher counter values correspond to events that occurred further in the past.

\noindent

\subsection{\textit{Software neuroscience model}}

\noindent \textbf{Simulating neural activity in hippocampal subregions.} 
Since simultaneous ensemble recording data from all subregions are not available at this time, a simple mechanistic model was built to simulate neural activity across all subregions during repeated exposure to a specific sequence of events (A, B, C, and D) in one of two contexts (Context1 or Context2). The temporal resolution of the simulation was 5 ms per timestep. Each event lasted 200 timesteps (1 s) and was separated by an interval of 1 s; each context lasted 1,700 timesteps (8.5 s). The number of neurons included across subregions followed proportions reported in the rodent brain (see Table 1). The first step in simulating the activity was to define the activity profiles of each subregion based on the literature, including the proportions of selective neurons \cite{mulders_neuron_1997, amaral_neurons_1990} and their mean firing rate curves during event and context presentations \cite{tsao_integrating_2018, allen2016Nonspatiala, shahbaba2022hippocampal, komorowski_robust_2009}. In the second step, spiking activities were generated using an inhomogeneous Poisson process given firing rate functions defined for each subregion, with stochastic noise added to simulate trial-to-trial variability (see Table 1). In total, spiking activity was generated through the network during 200 sequences of events (A, B, C, and D), with half the sequences presented in Context1 and half in Context2. We also simulated theta oscillations in CA1 as a sine wave, specifically six full theta cycles per event. The first cycle (40 timesteps) preceded the event and the remaining five cycles (200 timesteps) occurred during the event \cite{shahbaba2022hippocampal}. 

\noindent \textbf{Decoding the information contained in the ensemble activity.}
A multinomial logistic regression model using LASSO was implemented to decode (predict) the content represented in the simulated neural activity \cite{shahbaba2022hippocampal}, with separate models being used for each subregion. The model imposed an L1 penalty on the regression coefficients to eliminate neurons whose activity was not significantly associated with the classification task (e.g., event or context decoding) by shrinking their regression coefficient to zero.
The strength of the penalty---which determines the threshold for considering a neuron as significant---was set by 5-fold cross-validation. The model of each subregion was trained on the neural activity from 10 randomly selected sequences (including Events A, B, C, and D, as well as a “null” state) out of the 100 sequences in each context. The null state (500-ms period before the context) was added as an alternative for the model to select when the activity did not closely resemble any of the events (e.g., during intervals between events). Note that, to produce EC's separate event and context decodings (see Fig. \ref{fig:SimulationResults}a), EC also included a separate model focused on contextual representations which was trained using 20 randomly selected presentations of Context1 and Context2 (that included the sequence of four events). 

Before training the models, we computed the firing rate of each neuron during an optimal training window within event presentations (when activity likely reflected the current stimulus instead of past or future states; see Table 2) and z-normalized each neuron’s firing rate using its mean and standard deviation from the session. We then used those z-normalized firing rates to train the models. The optimal training windows were verified by comparing the classification accuracy of the validation set using a sliding window approach. For model testing, we predicted the decoding probabilities of the 5 states and 2 contexts for each subregion and used a Gaussian filter to smooth the probabilities. Note that model parameters were the same for all subregions, except for the subregion-specific parameters included in Tables 1 and 2. \\

\begingroup
\small

\setlength{\LTcapwidth}{\simtablewidth}
\setlength{\tabcolsep}{4pt}
\renewcommand{\arraystretch}{1.15}

\captionsetup[longtable]{
  font=customeighthalf,
  labelsep=pipe,
  justification=justified,
  singlelinecheck=false,
  width=\simtablewidth, 
  skip=8pt 
}

\begin{xltabular}{\simtablewidth}{
  @{} l *{4}{>{\centering\arraybackslash}X} @{}
}
\caption{\textbf{Anatomical and simulation parameters for each subregion.} Empirically derived anatomical population sizes \cite{mulders_neuron_1997, amaral_neurons_1990} (proportionately scaled in simulation) and statistical distributions used to define firing rates and additive noise (parenthetical values denote mean and variance).
}\label{sim_tab} \\
\toprule
\multirow{2.6}{*}{\textbf{Subregions}} & 
\multicolumn{1}{c}{\makecell{\textbf{Anatomical and} \\ \textbf{physiological evidence}}} &
\multicolumn{3}{c}{\textbf{Simulation parameters}} \\
\cmidrule(l){2-5}
& \textbf{Number of neurons} & \textbf{Number of neurons} & 
\textbf{Firing rate function} & \textbf{Noise distribution} \\
\midrule
\endfirsthead

\toprule
\multirow{2.6}{*}{\textbf{Subregions}} & 
\multicolumn{1}{c}{\makecell{\textbf{Anatomical and} \\ \textbf{physiological evidence}}} &
\multicolumn{3}{c}{\textbf{Simulation parameters}} \\
\cmidrule(l){2-5}
& \textbf{Number of neurons} & \textbf{Number of neurons} & 
\textbf{Firing rate distribution} & \textbf{Noise distribution} \\
\midrule
\endhead

\textbf{EC}   & $\sim$350,000     & 350    & Normal (.05, 10) & Normal (0, .05) \\
\textbf{DG}   & $\sim$1,000,000   & 1,000  & Gamma (2, 40)    & Normal (0, .01) \\
\textbf{CA3}  & $\sim$300,000     & 300    & Normal (.03, 1)  & Normal (0, .05) \\
\textbf{CA1}  & $\sim$400,000     & 400    & Normal (.01, 1)  & Normal (0, .05) \\

\bottomrule
\end{xltabular}

\endgroup

\begin{table}[h!]
\centering
\captionsetup{
  font=customeighthalf,
  labelsep=pipe,
  position=top,
  width=\decodingtablewidth,
  justification=justified,
  singlelinecheck=false,
  skip=8pt 
}
\caption{\textbf{Decoding hyperparameters for each subregion.} Training windows are reported as (start, end) in ms relative to training-class onset (t = 0), and Gaussian smoothing parameters are reported as standard deviation in ms. EC appears twice due to separate models being used to produce the event and context decodings.}\label{decoding_tab}
\renewcommand{\arraystretch}{1.3}
\setlength{\tabcolsep}{12pt} 
\begin{tabularx}{\decodingtablewidth}{
  @{} l *{3}{>{\centering\arraybackslash}X} @{}
}
\toprule
\textbf{Subregions} & 
\makecell{\textbf{Training}\\\textbf{class}} & 
\makecell{\textbf{Optimal training}\\\textbf{window (ms)}} & 
\makecell{\textbf{Std. dev. of}\\\textbf{Gaussian filter (ms)}} \\
\midrule
\textbf{EC}  & Context           & (0, 8500) & 100 \\
             & Event             & (50, 900)  & 20  \\
\textbf{DG}  & Event             & (200, 900)  & 15  \\
\textbf{CA3} & Events-in-context & (210, 260)  & 5   \\
\textbf{CA1} & Events-in-context & (275, 450)  & 8   \\
\bottomrule
\end{tabularx}
\end{table}

\normalsize

\subsection{Hardware model and verification}

\noindent \textbf{MATLAB Framework Implementation.}
During recall, the present context is compared against the Context AP-CAM using a hardware-accurate polynomial approximation function for analog proximity, followed by top-1 sorting to identify the best-matching context. Each input event is similarly compared against the Event AP-CAM, and the top-1 event is selected. The recurrent network consisting of event-to-event crossbar and context gating circuit then activates the connected events for the chosen context and event. CNTR-1 is initialized and compared with the MCAM array to align the timing of the present event, and its value is propagated to CNTR-2 and CNTR-3 to enable near-term and longer-term future predictions. These counters, operating at different rates, iteratively match against recurrent activations to identify upcoming events. Predicted events are tagged with analog voltages representing the present and two future events, while a ramp-voltage sweep latches timing flags in the CA1 module. Finally, a controlled voltage decay generates the temporal context signal, allowing integration across cycles to trace past events and support context-aware sequence prediction. The step-by-step operations of this hardware model are detailed in Algorithm 1.

\begin{algorithm}
\caption{HippocamPEX Hardware Model: Context-aware Sequence Prediction}
\begin{algorithmic}[1]
\State \textbf{Configuration}
\State \quad Generate context and event hypervectors (HVs)
\State \quad Load context AP-CAM and event AP-CAM with the corresponding HVs
\State \quad Load task-specific, software-generated activations into the recurrent heteroassociative crossbar 
\State \quad Load software-generated temporal information into MCAM array in CA3 for each context-event pair
\State \textbf{Input:} Present context hypervector $\mathbf{C}$ and event hypervectors $\mathbf{E}_{1:\tau}$
\State \textbf{Output:} Future context-aware event predictions ($\mathbf{E}_{\tau+1}, \mathbf{E}_{\tau+2}$), and temporal integration of  events in the sequence
\State Compute the analog proximity between the present context and the context AP-CAM using a hardware-mimicking polynomial approximation function.
\State Top-1 sorting to identify best matching context $\rightarrow \mathbf{C}^{S}$

\For{$t = 1$ to $\tau - 2$} 
    \State Compute the analog proximity between the present event $\mathbf{E}_t$ and the event AP-CAM.
    \State Top-1 sorting to identify best matching event $\rightarrow \mathbf{E}_t^{S}$
    \State Activate the connected events to the top-1 sorted event ($\mathbf{E}_t^{S}$) and context ($\mathbf{C}^{S}$) via recurrent network
    \State Initialize \texttt{CNTR-1} and compare against MCAM array to detect $\mathbf{E}_t^{S}$ timing
    \State Load \texttt{CNTR-1} value to \texttt{CNTR-2} and \texttt{CNTR-3} as initialization for the future predictions

    \While{No match for \texttt{CNTR-2} or \texttt{CNTR-3}}
        \If{Match(\texttt{CNTR-2}, activated events from the recurrent network)}
            \State Predict $\mathbf{E}_{t+1}$ 
        \EndIf
        \If{Match(\texttt{CNTR-3}, activated events from the recurrent network)}
            \State Predict $\mathbf{E}_{t+2}$ 
        \EndIf
        \State Increment \texttt{CNTR-2} \texttt{CNTR-3} 
    \EndWhile

    \State Tag voltages for $\mathbf{E}_t$ ($V^P$), $\mathbf{E}_{t+1}$ ($V^{F1}$), and $\mathbf{E}_{t+2}$ ($V^{F2}$)
    \For{$g = 0$ to $2^n-1$}
        \State Generate ramp voltage $V_{\text{ramp}} = V_{DD} - g \cdot \delta$
        \For{each event $e$}
            \If{$V_{\text{ramp}} < V_{\text{tag}}(e)$ and not previously triggered}
                \State Latch $g$ value to the sequencer flag register of $e$ in CA1
            \EndIf
        \EndFor
    \EndFor
    \State Decay all voltage tags by $v_{\text{decay}}$, reset future-1 and future-2 event tags
\EndFor

\end{algorithmic}
\end{algorithm}

Based on Algorithm 1, we implemented a MATLAB framework to model the full HippocamPEX hardware functionality. This model mimics the exact hardware behavior while enabling large-scale verification of context-aware sequence prediction. All core operations---including hypervector generation, event-context associations, recurrent activations, and temporal coding and integration---were faithfully incorporated, with the analog CAM’s polynomial approximation function used to replicate hardware proximity search. The verification process and associated results are illustrated in Figure \ref{fig_hardware_model_verification}.

\newpage
\section*{Acknowledgments}

This work was supported in part by the National Science Foundation under awards CCF-2319617, CCF-2319618, CCF-2319619, CAREER IOS-1150292, BCS-1439267, and DGE-1839285. This work was also partially supported by the National Institutes of Health under awards R01-MH115697, R01-DC017687, and R01-CA297869. The authors used AI for proofreading and shortening sections, not for content generation.


\newpage
\bibliography{sn-bibliography}

@inproceedings{stewart2012spaun,
  title={Spaun: A perception-cognition-action model using spiking neurons},
  author={Stewart, Terrence and Choo, Feng-Xuan and Eliasmith, Chris},
  booktitle={Proceedings of the Annual Meeting of the Cognitive Science Society},
  volume={34},
  number={34},
  year={2012}
}

@article{bouhadjar2023sequence,
  title={Sequence learning in a spiking neuronal network with memristive synapses},
  author={Bouhadjar, Younes and Siegel, Sebastian and Tetzlaff, Tom and Diesmann, Markus and Waser, Rainer and Wouters, Dirk J},
  journal={Neuromorphic Computing and Engineering},
  volume={3},
  number={3},
  pages={034014},
  year={2023},
  publisher={IOP Publishing}
}

@inproceedings{lindsey2022sequence,
  title={Sequence learning and consolidation on Loihi using on-chip plasticity},
  author={Lindsey, Jack and Aimone, James B},
  booktitle={Proceedings of the 2022 Annual Neuro-Inspired Computational Elements Conference},
  pages={70--72},
  year={2022}
}

@article{stan2024learning,
  title={Learning long sequences in spiking neural networks},
  author={Stan, Matei-Ioan and Rhodes, Oliver},
  journal={Scientific Reports},
  volume={14},
  number={1},
  pages={21957},
  year={2024},
  publisher={Nature Publishing Group UK London}
}

@inproceedings{siegel2023demonstration,
  title={Demonstration of neuromorphic sequence learning on a memristive array},
  author={Siegel, Sebastian and Ziegler, Tobias and Bouhadjar, Younes and Tetzlaff, Tom and Waser, Rainer and Dittmann, Regina and Wouters, Dirk},
  booktitle={Proceedings of the 2023 Annual Neuro-Inspired Computational Elements Conference},
  pages={108--114},
  year={2023}
}

@article{kreiser2018organizing,
  title={Organizing sequential memory in a neuromorphic device using dynamic neural fields},
  author={Kreiser, Raphaela and Aathmani, Dora and Qiao, Ning and Indiveri, Giacomo and Sandamirskaya, Yulia},
  journal={Frontiers in Neuroscience},
  volume={12},
  pages={717},
  year={2018},
  publisher={Frontiers Media SA}
}

@article{yin2023iris,
  title={{IRIS}: Integrated retinal functionality in image sensors},
  author={Yin, Zihan and Kaiser, Md Abdullah-Al and Camara, Lamine Ousmane and Camarena, Mark and Parsa, Maryam and Jacob, Ajey and Schwartz, Gregory and Jaiswal, Akhilesh},
  journal={Frontiers in Neuroscience},
  volume={17},
  pages={1241691},
  year={2023},
  publisher={Frontiers Media SA}
}

@article{yin2020xnor,
  title={{XNOR-SRAM}: In-memory computing {SRAM} macro for binary/ternary deep neural networks},
  author={Yin, Shihui and Jiang, Zhewei and Seo, Jae-Sun and Seok, Mingoo},
  journal={IEEE Journal of Solid-State Circuits},
  volume={55},
  number={6},
  pages={1733--1743},
  year={2020},
  publisher={IEEE}
}

@article{blundell2016model,
  title={Model-free episodic control},
  author={Blundell, Charles and Uria, Benigno and Pritzel, Alexander and Li, Yazhe and Ruderman, Avraham and Leibo, Joel Z and Rae, Jack and Wierstra, Daan and Hassabis, Demis},
  journal={arXiv preprint arXiv:1606.04460},
  year={2016}
}

@inproceedings{pritzel2017neural,
  title={Neural episodic control},
  author={Pritzel, Alexander and Uria, Benigno and Srinivasan, Sriram and Badia, Adria Puigdomenech and Vinyals, Oriol and Hassabis, Demis and Wierstra, Daan and Blundell, Charles},
  booktitle={International conference on machine learning},
  pages={2827--2836},
  year={2017},
  organization-hidden={PMLR}
}

@article{wayne2018unsupervised,
  title={Unsupervised predictive memory in a goal-directed agent},
  author={Wayne, Greg and Hung, Chia-Chun and Amos, David and Mirza, Mehdi and Ahuja, Arun and Grabska-Barwinska, Agnieszka and Rae, Jack and Mirowski, Piotr and Leibo, Joel Z and Santoro, Adam and others},
  journal={arXiv preprint arXiv:1803.10760},
  year={2018}
}

@article{le2021model,
  title={Model-based episodic memory induces dynamic hybrid controls},
  author={Le, Hung and Karimpanal George, Thommen and Abdolshah, Majid and Tran, Truyen and Venkatesh, Svetha},
  journal={Advances in Neural Information Processing Systems},
  volume={34},
  pages={30313--30325},
  year={2021}
}

@article{poursiami2025scalable,
  title={A scalable reinforcement learning framework inspired by hippocampal memory mechanisms for efficient contextual and sequential decision making},
  author={Poursiami, Hamed and Moshruba, Ayana and Cooper, Keiland W and Gobin, Derek and Kaiser, Md Abdullah-Al and Singh, Ankur and Noor, Rouhan and Shahbaba, Babak and Jaiswal, Akhilesh and Fortin, Norbert J and others},
  journal={Scientific Reports},
  volume={15},
  number={1},
  pages={25221},
  year={2025},
  publisher={Nature Publishing Group UK London}
}

@article{datta2022processing,
  title={A processing-in-pixel-in-memory paradigm for resource-constrained {TinyML} applications},
  author={Datta, Gourav and Kundu, Souvik and Yin, Zihan and Lakkireddy, Ravi Teja and Mathai, Joe and Jacob, Ajey P and Beerel, Peter A and Jaiswal, Akhilesh R},
  journal={Scientific Reports},
  volume={12},
  number={1},
  pages={14396},
  year={2022},
  publisher={Nature Publishing Group UK London}
}

@article{kaiser2025voltage,
  title={Voltage controlled magnetic tunnel junction based {ADC} less processing-in-pixel for extreme edge intelligence},
  author={Kaiser, Md Abdullah-Al and Datta, Gourav and Athas, Jordan and Duffee, Christian and Jacob, Ajey P and Amiri, Pedram Khalili and Beerel, Peter A and Jaiswal, Akhilesh R},
  journal={npj Unconventional Computing},
  volume={2},
  number={1},
  pages={14},
  year={2025},
  publisher={Nature Publishing Group UK London}
}

@article{ravichandran2026memristive,
  title={Memristive cellular neural networks for fast in-pixel computing},
  author={Ravichandran, Vignesh and Huang, Yi and Flannery, Bryce and Molom-Ochir, Tergel and Maurer, Tina and Asapu, Shiva and Abdel-Maksoud, Ali and Heermance, Nia and Yoo, Remy and Tackie, Joshua and others},
  journal={Nature Electronics},
  pages={1--11},
  year={2026},
  publisher={Nature Publishing Group UK London}
}

@article{zhang2024semantic,
  title={Semantic memory--based dynamic neural network using memristive ternary {CIM} and {CAM} for {2D} and {3D} vision},
  author={Zhang, Yue and Zhang, Woyu and Wang, Shaocong and Lin, Ning and Yu, Yifei and He, Yangu and Wang, Bo and Jiang, Hao and Lin, Peng and Xu, Xiaoxin and others},
  journal={Science Advances},
  volume={10},
  number={33},
  pages={eado1058},
  year={2024},
  publisher={American Association for the Advancement of Science}
}

@article{shahbaba2022hippocampal,
  title={Hippocampal ensembles represent sequential relationships among an extended sequence of nonspatial events},
  author={Shahbaba, Babak and Li, Lingge and Agostinelli, Forest and Saraf, Mansi and Cooper, Keiland W and Haghverdian, Derenik and Elias, Gabriel A and Baldi, Pierre and Fortin, Norbert J},
  journal={Nature Communications},
  volume={13},
  number={1},
  pages={787},
  year={2022},
  publisher={Nature Publishing Group UK London}
}

@article{aguirre2024hardware,
  title={Hardware implementation of memristor-based artificial neural networks},
  author={Aguirre, Fernando and Sebastian, Abu and Le Gallo, Manuel and Song, Wenhao and Wang, Tong and Yang, J Joshua and Lu, Wei and Chang, Meng-Fan and Ielmini, Daniele and Yang, Yuchao and others},
  journal={Nature Communications},
  volume={15},
  number={1},
  pages={1974},
  year={2024},
  publisher={Nature Publishing Group UK London}
}

@article{yan2024emerging,
  title={Emerging opportunities and challenges for the future of reservoir computing},
  author={Yan, Min and Huang, Can and Bienstman, Peter and Tino, Peter and Lin, Wei and Sun, Jie},
  journal={Nature Communications},
  volume={15},
  number={1},
  pages={2056},
  year={2024},
  publisher={Nature Publishing Group UK London}
}

@article{wang2025integrated,
  title={Integrated photonic recurrent processors},
  author={Wang, Yizhi and Luo, Lingzhi and Xing, Sizhe and Yao, Chunhui and Penty, Richard and Cheng, Qixiang},
  journal={Advanced Photonics},
  volume={7},
  number={5},
  pages={054003--054003},
  year={2025},
  publisher={Society of Photo-Optical Instrumentation Engineers}
}

@article{shen2023deep,
  title={Deep photonic reservoir computing recurrent network},
  author={Shen, Yi-Wei and Li, Rui-Qian and Liu, Guan-Ting and Yu, Jingyi and He, Xuming and Yi, Lilin and Wang, Cheng},
  journal={Optica},
  volume={10},
  number={12},
  pages={1745--1751},
  year={2023},
  publisher={Optica Publishing Group}
}

@article{allen2013evolution,
  title={The evolution of episodic memory},
  author={Allen, Timothy A and Fortin, Norbert J},
  journal={Proceedings of the National Academy of Sciences},
  volume={110},
  number={2},
  pages={10379--10386},
  year={2013},
  publisher={National Academy of Sciences}
}

@article{stark2018s,
  title={What’s in a context? Cautions, limitations, and potential paths forward},
  author={Stark, Shauna M and Reagh, Zachariah M and Yassa, Michael A and Stark, Craig EL},
  journal={Neuroscience Letters},
  volume={680},
  pages={77--87},
  year={2018},
  publisher={Elsevier}
}

@article{lisman1999relating,
  title={Relating hippocampal circuitry to function: recall of memory sequences by reciprocal dentate--{CA3} interactions},
  author={Lisman, John E},
  journal={Neuron},
  volume={22},
  number={2},
  pages={233--242},
  year={1999},
  publisher={Elsevier}
}

@article{polyn2008memory,
  title={Memory search and the neural representation of context},
  author={Polyn, Sean M and Kahana, Michael J},
  journal={Trends in Cognitive Sciences},
  volume={12},
  number={1},
  pages={24--30},
  year={2008},
  publisher={Elsevier}
}

@article{levy1996sequence,
  title={A sequence predicting {CA3} is a flexible associator that learns and uses context to solve hippocampal-like tasks},
  author={Levy, William B},
  journal={Hippocampus},
  volume={6},
  number={6},
  pages={579--590},
  year={1996},
  publisher={Wiley Online Library}
}

@article{buzsaki2018space,
  title={Space and time: the hippocampus as a sequence generator},
  author={Buzs{\'a}ki, Gy{\"o}rgy and Tingley, David},
  journal={Trends in Cognitive Sciences},
  volume={22},
  number={10},
  pages={853--869},
  year={2018},
  publisher={Elsevier}
}

@article{wikenheiser2015hippocampal,
  title={Hippocampal theta sequences reflect current goals},
  author={Wikenheiser, Andrew M and Redish, A David},
  journal={Nature Neuroscience},
  volume={18},
  number={2},
  pages={289--294},
  year={2015},
  publisher={Nature Publishing Group US New York}
}

@article{zhong2021dynamic,
  title={Dynamic memristor-based reservoir computing for high-efficiency temporal signal processing},
  author={Zhong, Yanan and Tang, Jianshi and Li, Xinyi and Gao, Bin and Qian, He and Wu, Huaqiang},
  journal={Nature Communications},
  volume={12},
  number={1},
  pages={408},
  year={2021},
  publisher={Nature Publishing Group UK London}
}

@article{zhong2022memristor,
  title={A memristor-based analogue reservoir computing system for real-time and power-efficient signal processing},
  author={Zhong, Yanan and Tang, Jianshi and Li, Xinyi and Liang, Xiangpeng and Liu, Zhengwu and Li, Yijun and Xi, Yue and Yao, Peng and Hao, Zhenqi and Gao, Bin and others},
  journal={Nature Electronics},
  volume={5},
  number={10},
  pages={672--681},
  year={2022},
  publisher={Nature Publishing Group UK London}
}

@article{eichenbaum_medial_2007,
    title = {The {Medial} {Temporal} {Lobe} and {Recognition} {Memory}},
    volume = {30},
    issn = {0147-006X, 1545-4126},
    url = {https://www.annualreviews.org/doi/10.1146/annurev.neuro.30.051606.094328},
    doi-hidden = {10.1146/annurev.neuro.30.051606.094328},
    language = {en},
    number = {1},
    urldate-hidden = {2025-06-13},
    journal = {Annual Review of Neuroscience},
    author = {Eichenbaum, H. and Yonelinas, A.P. and Ranganath, C.},
    month = jul,
    year = {2007},
    pages = {123--152},
}

@article{kumaran_what_2016,
    title = {What {Learning} {Systems} do {Intelligent} {Agents} {Need}? {Complementary} {Learning} {Systems} {Theory} {Updated}},
    volume = {20},
    issn = {13646613},
    shorttitle = {What {Learning} {Systems} do {Intelligent} {Agents} {Need}?},
    url = {https://linkinghub.elsevier.com/retrieve/pii/S1364661316300432},
    doi-hidden = {10.1016/j.tics.2016.05.004},
    language = {en},
    number = {7},
    urldate-hidden = {2025-02-28},
    journal = {Trends in Cognitive Sciences},
    author = {Kumaran, Dharshan and Hassabis, Demis and McClelland, James L.},
    month = jul,
    year = {2016},
    pages = {512--534},
}

@article{knierim_functional_2014,
    title = {Functional correlates of the lateral and medial entorhinal cortex: objects, path integration and local–global reference frames},
    volume = {369},
    shorttitle = {Functional correlates of the lateral and medial entorhinal cortex},
    url = {https://royalsocietypublishing.org/doi/10.1098/rstb.2013.0369},
    doi-hidden = {10.1098/rstb.2013.0369},
    number = {1635},
    urldate-hidden = {2025-02-28},
    journal = {Philosophical Transactions of the Royal Society B: Biological Sciences},
    publisher = {Royal Society},
    author = {Knierim, James J. and Neunuebel, Joshua P. and Deshmukh, Sachin S.},
    month = feb,
    year = {2014},
    pages = {20130369},
}

@article{mcclelland_why_1995,
    title = {Why there are complementary learning systems in the hippocampus and neocortex: {Insights} from the successes and failures of connectionist models of learning and memory},
    volume = {102},
    issn = {1939-1471},
    shorttitle = {Why there are complementary learning systems in the hippocampus and neocortex},
    doi-hidden = {10.1037/0033-295X.102.3.419},
    number = {3},
    journal = {Psychological Review},
    publisher = {American Psychological Association},
    author = {McClelland, James L. and McNaughton, Bruce L. and O'Reilly, Randall C.},
    year = {1995},
    note-hidden = {Place: US},
    pages = {419--457},
}

@article{vanstrien2009aAnatomy,
    title = {The anatomy of memory: an interactive overview of the parahippocampal–hippocampal network},
    volume = {10},
    copyright = {2009 Springer Nature Limited},
    issn = {1471-0048},
    shorttitle = {The anatomy of memory},
    url = {https://www.nature.com/articles/nrn2614},
    doi-hidden = {10.1038/nrn2614},
    language = {en},
    number = {4},
    urldate-hidden = {2025-02-28},
    journal = {Nature Reviews Neuroscience},
    publisher = {Nature Publishing Group},
    author = {{van Strien}, N. M. and Cappaert, N. L. M. and Witter, M. P.},
    month = apr,
    year = {2009},
    pages = {272--282},
}

@article{guzman_synaptic_2016,
    title = {Synaptic mechanisms of pattern completion in the hippocampal {CA3} network},
    volume = {353},
    url = {https://www.science.org/doi/10.1126/science.aaf1836},
    doi-hidden = {10.1126/science.aaf1836},
    number = {6304},
    urldate-hidden = {2025-02-28},
    journal = {Science},
    publisher = {American Association for the Advancement of Science},
    author = {Guzman, Segundo Jose and Schlögl, Alois and Frotscher, Michael and Jonas, Peter},
    month = sep,
    year = {2016},
    pages = {1117--1123},
}

@article{preston_interplay_2013,
    title = {Interplay of hippocampus and prefrontal cortex in memory},
    volume = {23},
    issn = {09609822},
    url = {http://dx.doi.org/10.1016/j.cub.2013.05.041},
    doi-hidden = {10.1016/j.cub.2013.05.041},
    number = {17},
    journal = {Current Biology},
    publisher = {Elsevier},
    author = {Preston, Alison R. and Eichenbaum, Howard},
    year = {2013},
    pages = {R764--R773},
}

@article{teyler2007Hippocampal,
    title = {The hippocampal indexing theory and episodic memory: {Updating} the index},
    volume = {17},
    issn = {10509631, 10981063},
    shorttitle = {The hippocampal indexing theory and episodic memory},
    url = {https://onlinelibrary.wiley.com/doi/10.1002/hipo.20350},
    doi-hidden = {10.1002/hipo.20350},
    language = {en},
    number = {12},
    urldate-hidden = {2022-05-10},
    journal = {Hippocampus},
    author = {Teyler, Timothy J. and Rudy, Jerry W.},
    month = dec,
    year = {2007},
    pages = {1158--1169},
}

@article{neunuebel2014ca3,
  title={{CA3} retrieves coherent representations from degraded input: direct evidence for {CA3} pattern completion and dentate gyrus pattern separation},
  author={Neunuebel, Joshua P and Knierim, James J},
  journal={Neuron},
  volume={81},
  number={2},
  pages={416--427},
  year={2014},
  publisher={Elsevier}
}

@article{okeefe1993Phaseb,
    title = {Phase relationship between hippocampal place units and the {EEG} theta rhythm},
    volume = {3},
    issn = {1098-1063},
    url = {https://onlinelibrary.wiley.com/doi/abs/10.1002/hipo.450030307},
    doi-hidden = {10.1002/hipo.450030307},
    language = {en},
    number = {3},
    urldate-hidden = {2025-03-18},
    journal = {Hippocampus},
    author = {O'Keefe, John and Recce, Michael L.},
    year = {1993},
    note-hidden = {\_eprint: https://onlinelibrary.wiley.com/doi/pdf/10.1002/hipo.450030307},
    pages = {317--330},
}

@article{johnson2007Neurala,
    chapter-hidden = {Articles},
    title = {Neural {Ensembles} in {CA3} {Transiently} {Encode} {Paths} {Forward} of the {Animal} at a {Decision} {Point}},
    volume = {27},
    copyright = {Copyright © 2007 Society for Neuroscience 0270-6474/07/2712176-14\$15.00/0},
    issn = {0270-6474, 1529-2401},
    url = {https://www.jneurosci.org/content/27/45/12176},
    doi-hidden = {10.1523/JNEUROSCI.3761-07.2007},
    language = {en},
    number = {45},
    urldate-hidden = {2025-06-13},
    journal = {Journal of Neuroscience},
    publisher = {Society for Neuroscience},
    author = {Johnson, Adam and Redish, A. David},
    month = nov,
    year = {2007},
    pages = {12176--12189},
}

@article{bornstein2017Reinstated,
    title = {Reinstated episodic context guides sampling-based decisions for reward},
    volume = {20},
    copyright = {2017 Nature Publishing Group, a division of Macmillan Publishers Limited. All Rights Reserved.},
    issn = {1546-1726},
    url = {https://www.nature.com/articles/nn.4573},
    doi-hidden = {10.1038/nn.4573},
    language = {en},
    number = {7},
    urldate-hidden = {2022-08-26},
    journal = {Nature Neuroscience},
    author = {Bornstein, Aaron M. and Norman, Kenneth A.},
    month = jul,
    year = {2017},
    pages = {997--1003},
}

@article{euston2012Rolea,
    title = {The {Role} of {Medial} {Prefrontal} {Cortex} in {Memory} and {Decision} {Making}},
    volume = {76},
    issn = {0896-6273},
    url = {https://www.sciencedirect.com/science/article/pii/S0896627312011087},
    doi-hidden = {10.1016/j.neuron.2012.12.002},
    number = {6},
    urldate-hidden = {2025-02-28},
    journal = {Neuron},
    author = {Euston, David R. and Gruber, Aaron J. and McNaughton, Bruce L.},
    month = dec,
    year = {2012},
    pages = {1057--1070},
}

@article{miller2001Integrativea,
    title = {An {Integrative} {Theory} of {Prefrontal} {Cortex} {Function}},
    volume = {24},
    issn = {0147-006X, 1545-4126},
    url = {https://www.annualreviews.org/content/journals/10.1146/annurev.neuro.24.1.167},
    doi-hidden = {10.1146/annurev.neuro.24.1.167},
    language = {en},
    number = {24},
    urldate-hidden = {2025-05-24},
    journal = {Annual Review of Neuroscience},
    publisher = {Annual Reviews},
    author = {Miller, Earl K. and Cohen, Jonathan D.},
    month = mar,
    year = {2001},
    pages = {167--202},
}

@article{rolls2013Mechanismsa,
    title = {The mechanisms for pattern completion and pattern separation in the hippocampus},
    volume = {7},
    issn = {1662-5137},
    url = {https://www.ncbi.nlm.nih.gov/pmc/articles/PMC3812781/},
    doi-hidden = {10.3389/fnsys.2013.00074},
    urldate-hidden = {2025-03-17},
    journal = {Frontiers in Systems Neuroscience},
    author = {Rolls, Edmund T.},
    month = oct,
    year = {2013},
    pages = {74},
}

@article{okeefe_dual_2005,
    title = {Dual phase and rate coding in hippocampal place cells: theoretical significance and relationship to entorhinal grid cells},
    volume = {15},
    issn = {1050-9631},
    shorttitle = {Dual phase and rate coding in hippocampal place cells},
    doi-hidden = {10.1002/hipo.20115},
    language = {eng},
    number = {7},
    journal = {Hippocampus},
    author = {O'Keefe, John and Burgess, Neil},
    year = {2005},
    pages = {853--866},
}

@article{qasim_phase_2021,
    title = {Phase precession in the human hippocampus and entorhinal cortex},
    volume = {184},
    issn = {0092-8674},
    url = {https://www.sciencedirect.com/science/article/pii/S0092867421004967},
    doi-hidden = {10.1016/j.cell.2021.04.017},
    number = {12},
    urldate-hidden = {2026-09-02},
    journal = {Cell},
    author = {Qasim, Salman E. and Fried, Itzhak and Jacobs, Joshua},
    month = jun,
    year = {2021},
    pages = {3242--3255},
}

@article{borzello_assessments_2023,
    title = {Assessments of dentate gyrus function: discoveries and debates},
    volume = {24},
    copyright = {2023 This is a U.S. Government work and not under copyright protection in the US; foreign copyright protection may apply},
    issn = {1471-0048},
    shorttitle = {Assessments of dentate gyrus function},
    url = {https://www.nature.com/articles/s41583-023-00710-z},
    doi-hidden = {10.1038/s41583-023-00710-z},
    language = {en},
    number = {8},
    urldate-hidden = {2026-09-02},
    journal = {Nature Reviews Neuroscience},
    publisher = {Nature Publishing Group},
    author = {Borzello, Mia and Ramirez, Steve and Treves, Alessandro and Lee, Inah and Scharfman, Helen and Stark, Craig and Knierim, James J. and Rangel, Lara M.},
    month = aug,
    year = {2023},
    pages = {502--517},
}

@article{mulders_neuron_1997,
    title = {Neuron numbers in the presubiculum, parasubiculum, and entorhinal area of the rat},
    volume = {385},
    issn = {0021-9967},
    language = {eng},
    number = {1},
    journal = {The Journal of Comparative Neurology},
    author = {Mulders, W. H. and West, M. J. and Slomianka, L.},
    month = aug,
    year = {1997},
    pages = {83--94},
}

@article{amaral_neurons_1990,
    year = {1990}, 
    month = {12}, 
    title = {Neurons, numbers and the hippocampal network}, 
    author = {Amaral, David G. and Ishizuka, Norio and Claiborne, Brenda}, 
    journal = {Progress in Brain Research}, 
    issn = {0079-6123}, 
    doi-hidden = {10.1016/S0079-6123(08)61237-6}, 
    pmid = {2203093}, 
    pages = {1--11}, 
    volume = {83}
}

@article{tsao_integrating_2018,
    title = {Integrating time from experience in the lateral entorhinal cortex},
    volume = {561},
    copyright = {2018 Springer Nature Limited},
    issn = {1476-4687},
    url = {https://www.nature.com/articles/s41586-018-0459-6},
    doi-hidden = {10.1038/s41586-018-0459-6},
    language = {en},
    number = {7721},
    urldate-hidden = {2026-09-04},
    journal = {Nature},
    publisher = {Nature Publishing Group},
    author = {Tsao, Albert and Sugar, Jørgen and Lu, Li and Wang, Cheng and Knierim, James J. and Moser, May-Britt and Moser, Edvard I.},
    month = sep,
    year = {2018},
    pages = {57--62},
}

@article{howard_distributed_2002,
    title = {A {Distributed} {Representation} of {Temporal} {Context}},
    volume = {46},
    issn = {0022-2496},
    url = {https://www.sciencedirect.com/science/article/pii/S0022249601913884},
    doi-hidden = {10.1006/jmps.2001.1388},
    number = {3},
    urldate-hidden = {2026-09-04},
    journal = {Journal of Mathematical Psychology},
    author = {Howard, Marc W. and Kahana, Michael J.},
    month = jun,
    year = {2002},
    pages = {269--299},
}

@article{shankar_timing_2010,
    series = {Computational {Cognitive} {Neuroscience} {III}: {Selected} presentations from {CCNC}-09},
    title = {Timing using temporal context},
    volume = {1365},
    issn = {0006-8993},
    url = {https://www.sciencedirect.com/science/article/pii/S0006899310016355},
    doi-hidden = {10.1016/j.brainres.2010.07.045},
    urldate-hidden = {2026-09-04},
    journal = {Brain Research},
    author = {Shankar, Karthik H. and Howard, Marc W.},
    month = dec,
    year = {2010},
    pages = {3--17},
}

@article{lavenex_hippocampal-neocortical_2000,
    title = {Hippocampal-neocortical interaction: {A} hierarchy of associativity},
    volume = {10},
    issn = {1098-1063},
    shorttitle = {Hippocampal-neocortical interaction},
    url = {https://onlinelibrary.wiley.com/doi/abs/10.1002/1098-1063%282000%2910%3A4%3C420%3A%3AAID-HIPO8%3E3.0.CO%3B2-5},
    doi-hidden = {10.1002/1098-1063(2000)10:4<420::AID-HIPO8>3.0.CO;2-5},
    language = {en},
    number = {4},
    urldate-hidden = {2026-09-04},
    journal = {Hippocampus},
    author = {Lavenex, Pierre and Amaral, David G.},
    year = {2000},
    pages = {420--430},
}

@article{furtak_functional_2007,
    title = {Functional neuroanatomy of the parahippocampal region in the rat: {The} perirhinal and postrhinal cortices},
    volume = {17},
    issn = {1098-1063},
    shorttitle = {Functional neuroanatomy of the parahippocampal region in the rat},
    url = {https://onlinelibrary.wiley.com/doi/abs/10.1002/hipo.20314},
    doi-hidden = {10.1002/hipo.20314},
    language = {en},
    number = {9},
    urldate-hidden = {2026-09-04},
    journal = {Hippocampus},
    author = {Furtak, Sharon C. and Wei, Shau-Ming and Agster, Kara L. and Burwell, Rebecca D.},
    year = {2007},
    pages = {709--722},
}

@article{kerr_functional_2007,
    title = {Functional neuroanatomy of the parahippocampal region: {The} lateral and medial entorhinal areas},
    volume = {17},
    issn = {1098-1063},
    shorttitle = {Functional neuroanatomy of the parahippocampal region},
    url = {https://onlinelibrary.wiley.com/doi/abs/10.1002/hipo.20315},
    doi-hidden = {10.1002/hipo.20315},
    language = {en},
    number = {9},
    urldate-hidden = {2026-09-04},
    journal = {Hippocampus},
    author = {Kerr, Kristin M. and Agster, Kara L. and Furtak, Sharon C. and Burwell, Rebecca D.},
    year = {2007},
    pages = {697--708},
}

@article{allen2016Nonspatiala,
    title = {Nonspatial sequence coding in {CA1} neurons},
    volume = {36},
    issn = {15292401},
    doi-hidden = {10.1523/JNEUROSCI.2874-15.2016},
    number = {5},
    journal = {Journal of Neuroscience},
    author = {Allen, Timothy A. and Salz, Daniel M. and McKenzie, Sam and Fortin, Norbert J.},
    year = {2016},
    pages = {1547--1563},
}

@article{komorowski_robust_2009,
    chapter-hidden = {Articles},
    title = {Robust {Conjunctive} {Item}–{Place} {Coding} by {Hippocampal} {Neurons} {Parallels} {Learning} {What} {Happens} {Where}},
    volume = {29},
    copyright = {Copyright © 2009 Society for Neuroscience 0270-6474/09/299918-12\$15.00/0},
    issn = {0270-6474, 1529-2401},
    url = {https://www.jneurosci.org/content/29/31/9918},
    doi-hidden = {10.1523/JNEUROSCI.1378-09.2009},
    language = {en},
    number = {31},
    urldate-hidden = {2026-09-04},
    journal = {Journal of Neuroscience},
    publisher = {Society for Neuroscience},
    author = {Komorowski, Robert W. and Manns, Joseph R. and Eichenbaum, Howard},
    month = aug,
    year = {2009},
    pages = {9918--9929},
}

@article{damphousse_vicarious_2026,
    title = {The vicarious nature of hippocampal theta sequences},
    volume = {381},
    issn = {0962-8436},
    url = {https://doi.org/10.1098/rstb.2025.0242},
    doi-hidden = {10.1098/rstb.2025.0242},
    number = {1954},
    urldate-hidden = {2026-09-04},
    journal = {Philosophical Transactions of the Royal Society B: Biological Sciences},
    author = {Damphousse, Chelsey C. and Gagliardi, Celia M. and Peterson, Jen G. and Feng, Christina and Schmidt, Brandy and Mugan, Ugurcan and Redish, A. David},
    month = jul,
    year = {2026},
    pages = {20250242},
}

@article{whittington_tolman-eichenbaum_2020,
    title = {The {Tolman}-{Eichenbaum} {Machine}: {Unifying} {Space} and {Relational} {Memory} through {Generalization} in the {Hippocampal} {Formation}},
    volume = {183},
    issn = {0092-8674},
    shorttitle = {The {Tolman}-{Eichenbaum} {Machine}},
    url = {https://www.sciencedirect.com/science/article/pii/S009286742031388X},
    doi-hidden = {10.1016/j.cell.2020.10.024},
    number = {5},
    urldate-hidden = {2026-09-04},
    journal = {Cell},
    author = {Whittington, James C. R. and Muller, Timothy H. and Mark, Shirley and Chen, Guifen and Barry, Caswell and Burgess, Neil and Behrens, Timothy E. J.},
    month = nov,
    year = {2020},
    pages = {1249--1263},
}

@article{eichenbaum1999Hippocampus,
    title = {The Hippocampus, Memory, and Place Cells: Is It Spatial Memory or a Memory Space ?},
    volume = {23},
    journal = {Neuron},
    author = {Eichenbaum, Howard and Dudchenko, Paul and Wood, Emma and Shapiro, Matthew and Tanila, Heikki},
    year = {1999},
    pages = {209--226},
}

@article{brown_neural_2026,
    title = {Neural circuit models for evidence accumulation through choice-selective sequences},
    volume = {17},
    copyright = {2026 The Author(s)},
    issn = {2041-1723},
    url = {https://www.nature.com/articles/s41467-026-70267-9},
    doi-hidden = {10.1038/s41467-026-70267-9},
    language = {en},
    number = {1},
    urldate-hidden = {2026-09-04},
    journal = {Nature Communications},
    publisher = {Nature Publishing Group},
    author = {Brown, Lindsey S. and Cho, Jounhong Ryan and Bolkan, Scott S. and Nieh, Edward H. and Schottdorf, Manuel and Tank, David W. and Brody, Carlos D. and Witten, Ilana B. and Goldman, Mark S.},
    month = mar,
    year = {2026},
    pages = {4055},
}

@article{maren_contextual_2013,
    title = {The contextual brain: implications for fear conditioning, extinction and psychopathology},
    volume = {14},
    copyright = {2013 Springer Nature Limited},
    issn = {1471-0048},
    shorttitle = {The contextual brain},
    url = {https://www.nature.com/articles/nrn3492},
    doi-hidden = {10.1038/nrn3492},
    language = {en},
    number = {6},
    urldate-hidden = {2026-09-04},
    journal = {Nature Reviews Neuroscience},
    publisher = {Nature Publishing Group},
    author = {Maren, Stephen and Phan, K. Luan and Liberzon, Israel},
    month = jun,
    year = {2013},
    pages = {417--428},
}

@article{bar_proactive_2009,
    title = {The proactive brain: memory for predictions},
    volume = {364},
    issn = {0962-8436},
    shorttitle = {The proactive brain},
    url = {https://doi.org/10.1098/rstb.2008.0310},
    doi-hidden = {10.1098/rstb.2008.0310},
    number = {1521},
    urldate-hidden = {2026-09-04},
    journal = {Philosophical Transactions of the Royal Society B: Biological Sciences},
    author = {Bar, Moshe},
    month = may,
    year = {2009},
    pages = {1235--1243},
}

@article{carr_hippocampal_2011,
    title = {Hippocampal replay in the awake state: a potential substrate for memory consolidation and retrieval},
    volume = {14},
    copyright = {2011 Springer Nature America, Inc.},
    issn = {1546-1726},
    shorttitle = {Hippocampal replay in the awake state},
    url = {https://www.nature.com/articles/nn.2732},
    doi-hidden = {10.1038/nn.2732},
    language = {en},
    number = {2},
    urldate-hidden = {2026-09-04},
    journal = {Nature Neuroscience},
    publisher = {Nature Publishing Group},
    author = {Carr, Margaret F. and Jadhav, Shantanu P. and Frank, Loren M.},
    month = feb,
    year = {2011},
    pages = {147--153},
}

@article{foster_replay_2017,
    title = {Replay {Comes} of {Age}},
    volume = {40},
    issn = {1545-4126},
    url = {https://www.annualreviews.org/content/journals/10.1146/annurev-neuro-072116-031538},
    doi-hidden = {10.1146/annurev-neuro-072116-031538},
    urldate-hidden = {2026-09-04},
    journal = {Annual Review of Neuroscience},
    publisher = {Annual Reviews},
    author = {Foster, David J.},
    year = {2017},
    pages = {581--602},
}


\end{document}